\documentclass[aps,prl,reprint,superscriptaddress]{revtex4-2}

\usepackage{float}
\usepackage{stmaryrd}
\usepackage{gensymb}
\usepackage{textcomp}
\usepackage{graphicx}
\usepackage{amsmath}
\usepackage{amssymb}
\usepackage[colorinlistoftodos]{todonotes}
\usepackage{siunitx}
\usepackage[absolute,overlay]{textpos}
\usepackage[colorlinks=true, citecolor={black}, urlcolor={blue!60!black}, linkcolor = {black}]{hyperref}
\usepackage{comment}
\usepackage{bold-extra}

\makeatletter

\newcommand{\fmarki}{*}
\newcommand{\fmarkii}{\ensuremath{\dagger}}
\newcommand{\fmarkiii}{\ensuremath{\ddagger}}
\newcommand{\fmarkiv}{\ensuremath{\mathsection}}
\newcommand{\fmarkv}{\ensuremath{\mathparagraph}}
\newcommand{\fmarkvi}{\ensuremath{\|}}
\newcommand{\fmarkvii}{**}
\newcommand{\fmarkviii}{\ensuremath{\dagger\dagger}}
\newcommand{\fmarkix}{\ensuremath{\ddagger\ddagger}}
\def\@fnsymbol#1{{\ifcase#1\or \fmarki\or \fmarkii\or \fmarkiii\or \fmarkiv\or \fmarkv\or \fmarkvi\or \fmarkvii\or \fmarkviii\or \fmarkix \else\@ctrerr\fi}}
\makeatother

\renewcommand{\fmarki}{\ensuremath{\dagger}}
\renewcommand{\fmarkii}{*}

\def\uppi~{$\mathrm{\pi}$}

\newcommand{\fl}[1]{\textbf{(#1)}}
\newcommand{\mui}[1]{$\mathrm{\mu}_{\mathrm{#1}}$}
\newcommand{\var}[2]{$#1_{\mathrm{#2}}$}

\newcommand{\phiD}{$\phi_{\mathrm{\Delta}}$}
\newcommand{\corrG}[1]{$G_{\mathrm{corr}}^{\mathrm{#1}}$}
\newcommand{\ABS}[1]{$V_{\mathrm{ABS}}^{\mathrm{#1}}$}

\begin{document}

\title{Observation of edge and bulk states in a three-site Kitaev chain}

\author{Sebastiaan L. D. ten Haaf}
\affiliation{QuTech and Kavli Institute of Nanoscience, Delft University of Technology, Delft, 2600 GA, The Netherlands}

\author{Yining Zhang}
\affiliation{QuTech and Kavli Institute of Nanoscience, Delft University of Technology, Delft, 2600 GA, The Netherlands}

\author{Qingzhen Wang}
\affiliation{QuTech and Kavli Institute of Nanoscience, Delft University of Technology, Delft, 2600 GA, The Netherlands}

\author{Alberto Bordin}
\affiliation{QuTech and Kavli Institute of Nanoscience, Delft University of Technology, Delft, 2600 GA, The Netherlands}

\author{Chun-Xiao Liu}
\affiliation{QuTech and Kavli Institute of Nanoscience, Delft University of Technology, Delft, 2600 GA, The Netherlands}

\author{Ivan Kulesh}
\affiliation{QuTech and Kavli Institute of Nanoscience, Delft University of Technology, Delft, 2600 GA, The Netherlands}

\author{Vincent P. M. Sietses}
\affiliation{QuTech and Kavli Institute of Nanoscience, Delft University of Technology, Delft, 2600 GA, The Netherlands}

\author{Christian G. Prosko}
\affiliation{QuTech and Kavli Institute of Nanoscience, Delft University of Technology, Delft, 2600 GA, The Netherlands}

\author{Di~Xiao}
\affiliation{Department of Physics and Astronomy, Purdue University, West Lafayette, 47907, Indiana, USA}

\author{Candice Thomas}
\affiliation{Department of Physics and Astronomy, Purdue University, West Lafayette, 47907, Indiana, USA}

\author{Michael J. Manfra}
\affiliation{Department of Physics and Astronomy, Purdue University, West Lafayette, 47907, Indiana, USA}
\affiliation{Elmore School of Electrical and Computer Engineering, ~Purdue University, West Lafayette, 47907, Indiana, USA}
\affiliation{School of Materials Engineering, Purdue University, West Lafayette, 47907, Indiana, USA}

\author{Michael Wimmer}
\affiliation{QuTech and Kavli Institute of Nanoscience, Delft University of Technology, Delft, 2600 GA, The Netherlands}

\author{Srijit Goswami}\email{s.goswami@tudelft.nl}
\affiliation{QuTech and Kavli Institute of Nanoscience, Delft University of Technology, Delft, 2600 GA, The Netherlands}
\begin{abstract}
A chain of quantum dots (QDs) in semiconductor-superconductor hybrid systems can form an artificial Kitaev chain hosting Majorana bound states (MBSs)~\cite{DasSarma2012,Leijnse2012,Fulga2013}.
These zero-energy states are expected to be localised on the edges of the chain~\cite{Kitaev2001}, at the outermost QDs.
The remaining QDs, comprising the bulk, are predicted to host an excitation gap that protects the MBSs at the edges from local on-site perturbations. 
Here we demonstrate this connection between the bulk and edges in a minimal system, by engineering a three-site Kitaev chain in a two-dimensional electron gas.
Through direct tunneling spectroscopy on each site, we show that the appearance of stable zero-bias conductance peaks at the outer QDs is correlated with the presence of an excitation gap in the middle QD.
Furthermore, we show that this gap can be controlled by applying a superconducting phase difference between the two hybrid segments and that the MBSs are robust only when the excitation gap is present.
We find a close agreement between experiments and the original Kitaev model, thus confirming key predictions for MBSs in a three-site chain. 
\end{abstract}

\maketitle
\section*{Introduction}
The study of topology in condensed matter has generated interest in engineering quantum phases hosting modes that are robust to external perturbations~\cite{Kane2005_QSHGraphene, Kane2010_topo_ins_review, Liang2011_topo_review, Alicea2012}.
In particular, realisations of 1-D topological superconductors~\cite{Lutchyn_PRL_2010, Oreg_PRL_2010} are expected to host robust edge modes known as Majorana bound states (MBSs), first predicted by the Kitaev chain model~\cite{Kitaev2001}.
Such states are separated by a bulk region with an excitation gap that prevents local perturbations from affecting the zero-energy modes at the edges, a consequence of the so-called bulk-edge correspondence~\cite{Halperin1982_edgestatesHall, Wen1995_topo_order}.
Experimentally, much effort has gone towards top-down approaches to engineer such systems by coupling semiconductors to superconductors~\cite{Prada2020}.
However, microscopic disorder in these systems complicates the study of MBSs~\cite{Liu2017Andreev, Reeg2018_ABSpinning, Vuik_quasi_Majorana_disorder,Pan_Physical_mechanism_PRR_2020,Sanker2021_disorder,kouwenhoven2024perspectivemajoranaboundstateshybrid}.
Alternatively, bottom-up approaches aim to mitigate this, e.g. by constructing a Kitaev chain atom-by-atom~\cite{Jian2014_topo_atom_chain,NadjPerge2014_atom_chain} or by engineering arrays of quantum dots (QDs) in semiconductor-superconductor hybrids~\cite{DasSarma2012, Fulga2013}.
An implementation of the latter approach with two QDs~\cite{Leijnse2012} was recently demonstrated in nanowires~\cite{Dvir2023} and two-dimensional electron gases (2DEGs)~\cite{tenHaaf2024}. 
However, in a two-site chain neither QD site can be associated with a bulk. 
In contrast, a chain with three QDs constitutes a minimal system where one could distinguish distinct edges hosting MBSs (the outer QDs) and a bulk (the middle QD).
Recent work demonstrated stable zero-energy modes in such a system~\cite{Bordin2024threesite}, but was unable to investigate the density of states of the middle QD.

In this work, we realise a three-site Kitaev chain in an InSbAs 2DEG and perform a detailed study of the bulk and edge states in the system. 
Crucially, ohmic contacts attached to each QD allow us to directly probe the density of states at all three sites. 
By controlling the interdot couplings and the superconducting phase difference, we can tune the system such that robust zero-bias conductance peaks (ZBPs) arise on the outer QDs.  
We demonstrate that these correlated ZBPs at the edges are accompanied by an excitation gap in the middle QD.
This gap can be controlled by tuning the phase difference between the superconductors, allowing us to establish a clear correlation between the presence of isolated zero-energy edge modes and their robustness against on-site perturbations. 
In particular, access to every QD allows us to track how the weight of the Majorana wavefunction evolves across each site as the device parameters are varied. 
Finally, we construct a phase diagram of a finite Kitaev chain by identifying regions in parameter space where ZBPs are observed. 
Through continuous control over the interdot couplings and the QD electrochemical potential energies, we show that this region grows when extending from a two-site chain to a three-site chain. 
These findings are in close agreement with the original Kitaev model, and provide important insights for more advanced experiments with MBSs. 

\begin{figure*}[t!]
\centering
    \includegraphics[width = \textwidth]{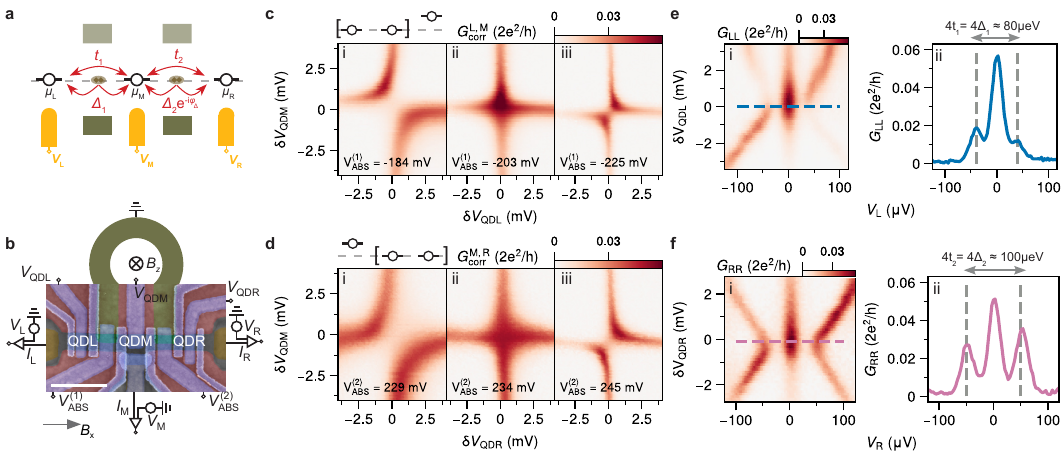}
    \caption{\textbf{Device, model and characterisation of two-site pairs.} \fl{a} Energy level diagram of the three-site Kitaev chain model, indicating the different interdot couplings in the system.
    \fl{b} Scanning electron micrograph of the device. 
    The scale bar is \SI{500}{\nano\meter}.
    Positions of gate-defined QDs are indicated. 
    Two superconducting strips are connected in a loop with a diameter of \SI{10}{\micro \meter} and kept grounded (not drawn to scale). The DC circuit diagram shows a four-terminal measurement set-up (a full circuit diagram, including resonators for reflectometry measurements, is shown in Fig.~S1).
    \fl{c} Measured correlated conductance \corrG{L,M}$=\sqrt{G_{\mathrm{LL}}\cdot G_{\mathrm{MM}}}$ as a function of \var{V}{QDL} and \var{V}{QDM}, upon varying \ABS{(1)}. 
    The notation $\delta V_{\mathrm{QDi}}$ refers to the voltage set with respect to the nearest charge degeneracy point. 
    \fl{d} Measured correlated conductance \corrG{M,R}$=\sqrt{G_{\mathrm{MM}}\cdot G_{\mathrm{RR}}}$ as a function of \var{V}{QDR} and \var{V}{QDM}, upon varying \ABS{(2)}.  
    The disappearance of the avoided crossing in panels (c.ii) and (d.ii) signifies satisfying the two-site sweet spot conditions.
    \fl{e} Finite bias spectroscopy measurement of \var{G}{LL} at the left QD pair sweet spot, while varying \var{V}{QDL}. 
    \fl{f} Finite bias spectroscopy of \var{G}{RR} at the right QD pair sweet spot, while varying \var{V}{QDR}.
    Line-traces taken at the minima of the higher energy excitations allow for estimating the experimental coupling amplitudes at the sweet spots (panels e,f.ii). 
    Results presented use the $\downarrow\uparrow\uparrow$ spin-configuration.
    Further characterisation of the two-site pairs is presented in Fig.~S4.
   }
    \label{fig:Fig1}
\end{figure*}
\begin{figure*}
\centering
    \includegraphics[width = \textwidth]{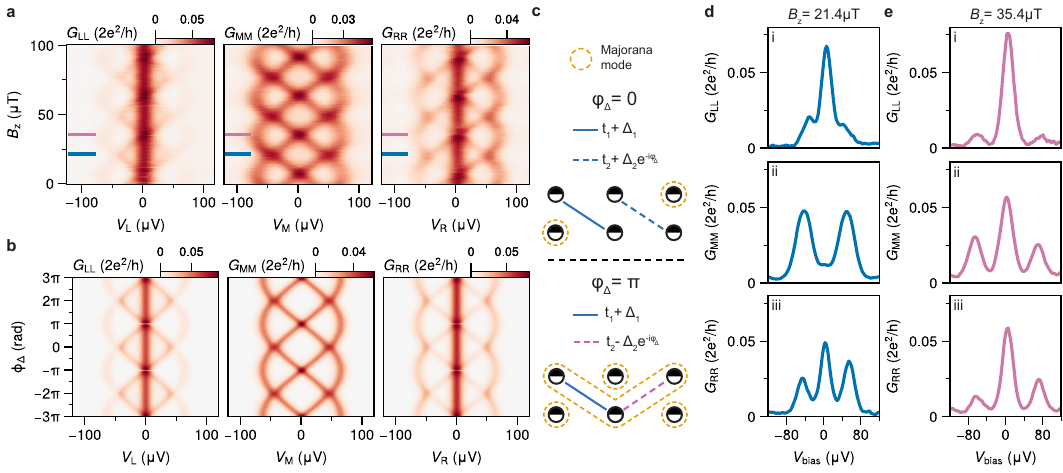}
    \caption{\textbf{Phase control at the three-site sweet spot.} 
    \fl{a} Measured conductances \var{G}{LL}, \var{G}{MM} and \var{G}{RR} at the sweet spot configuration obtained in Fig~1, as a function of magnetic field \var{B}{z} applied perpendicular to the superconducting loop.
    When either of the outer QDs is set off-resonance, \var{B}{z} no longer affects the conductance spectra (shown in Fig.~S4), indicating the observed behaviour here is an effect on the full three-site chain.
    \fl{b} Evolution of numerically calculated conductances in a three-site Kitaev chain upon varying \var{\phi}{\Delta}. 
    Numerical parameters used are \mui{L}~=~\mui{M}~=~\mui{R}~=~0, \var{t}{1} = \var{\Delta}{1}~=~\SI{20}{\micro \electronvolt} and \var{t}{2} = \var{\Delta}{2}~=~\SI{25}{\micro \electronvolt} (matching estimations in Fig.~1e,f). 
    \fl{c} Visual representation of the three-site Kitaev chain Hamiltonian at \var{\phi}{\Delta}~=~0 and \var{\phi}{\Delta}~=~\uppi~, in the Majorana basis. 
    Connecting lines indicate non-zero coupling terms between Majoranas for each case. Dashed circles highlight the distribution of uncoupled Majorana modes (further detailed in Methods). 
    \fl{d} Line-traces taken from (a) at $B_{z}$~=~\SI{21.4}{\micro \tesla}, corresponding to \var{\phi}{\Delta}~=~0. 
    The finite conductance at zero bias in 2d.ii can be largely attributed to thermal broadening, investigated in Fig.~S5.
    \fl{e} Line-traces taken at $B_{z}$~=~\SI{35.4}{\micro \tesla}, corresponding to \var{\phi}{\Delta}~=~\uppi~.}
    \label{fig:Fig2}
\end{figure*}

\subsection*{The Kitaev model for a three-site QD chain}
Implementing a Kitaev chain on a QD array requires control over the electrochemical potential energy of each site, and the amplitudes of couplings between neighbouring sites. 
This can be engineered by coupling spin-polarised QDs via Andreev bound states in semiconductor-superconductor hybrids~\cite{Dvir2023}.
Here, a hopping interaction ($t$) occurs through elastic co-tunneling (ECT).
Crossed Andreev reflection (CAR) provides a pairing interaction ($\Delta$) via the creation or breaking of Cooper pairs in the superconductor~\cite{Liu2022, Bordin2022_ect_car,Bordin2024_ectcar, Liu2024_ectcar}.
Intrinsic spin-orbit coupling allows for CAR between neighbouring sites with the same spin, or ECT between neighbours with opposite spin, otherwise forbidden due to spin-conservation~\cite{Wang2022a}. 
When the Zeeman energy in each QD is large (\var{E}{z}~$\gg$~\var{t}{},~\var{\Delta}{}), a single spin-species dominates transport at charge degeneracy points, thus emulating the Kitaev model~\cite{Tsintzis2022,luethi2024}.
To distinguish different charge configurations for the three QDs, the spin configuration is used as a label.
The energy level diagram for a chain with three QDs is shown in Fig.~1a.
Electrochemical potential energies are denoted \var{\mu}{i}, where $i\in \{\mathrm{L},\mathrm{M},\mathrm{R}\}$ refers to the left, middle or right QD respectively.
The effective couplings between the left and middle QD, \var{t}{1} and \var{\Delta}{1}, and between the middle and right QD, \var{t}{2} and \var{\Delta}{2}, are indicated.
Uncoupled MBSs arise on the outer QDs at a so-called `sweet spot' in parameter space, when all three QDs are aligned with the Fermi level of the superconductor (\var{\mu}{i}~=~0) and the interdot couplings are equal in amplitude pairwise ($\lvert t_1\rvert=\lvert\Delta_1\rvert$, $\lvert t_2\rvert=\lvert\Delta_2\rvert$).
Notably, the coupling amplitudes are not necessarily purely real in a system with more than two QDs~\cite{Fulga2013, Bordin2024threesite, Liu2024scaling}. 
A gauge degree of freedom allows one to take \var{t}{1}, \var{t}{2} and \var{\Delta}{1} to be real and assign a complex phase only to \var{\Delta}{2}, denoted \var{\phi}{\Delta}.
We compare our experimental results to this three-site Kitaev model. 
Numerical simulations of the conductance are performed using a scattering matrix approach with experimentally extracted parameters.
A more detailed description is included in Methods.

\subsection*{Device and measurement set-up}
A scanning electron micrograph of the measured device is shown in Fig.~1b. 
Large gates (red) define a quasi-1D channel across two thin aluminium strips that are connected by a continuous loop.
An external magnetic field \var{B}{z} applied perpendicular to the loop controls the superconducting phase difference \var{\phi}{SC} between the two strips, with a flux period of \SI{28}{\micro T} (Fig.~S1). 
The energy of the ABSs in the left and right hybrid regions can be tuned by voltages \ABS{(1)} and \ABS{(2)} respectively.
Narrow gates define three QDs in the channel.
Their electrochemical potentials are controlled by voltages \var{V}{QDL}, \var{V}{QDM} and \var{V}{QDR} respectively.
Voltages applied to a left, middle and right lead (\var{V}{L}, \var{V}{M} and \var{V}{R}) can be varied and the currents in each lead (\var{I}{L}, \var{I}{M} and \var{I}{R}) can be measured independently. 
Lock-in amplifiers allow for direct measurements of the local conductance at each probe ($G_{ii}=\frac{dI_{i}}{dV_{i}}$,  $i \in \{\mathrm{L},\mathrm{M},\mathrm{R}\}$).
To capture features appearing simultaneously on multiple sites, the correlated conductance $G_{\mathrm{corr}}^{i,j}$=$\sqrt{G_{ii}\cdot G_{jj}}$ is extracted when relevant.
Results in the main text are obtained using a single orbital in each QD, characterised in Fig.~S2, S3. 
All measurements are performed at \var{B}{x}~=~\SI{200}{\milli \tesla}, applied perpendicular to the spin-orbit field in these systems~\cite{WangQ2022}.
\newline
\begin{figure*}[t!]
\centering
    \includegraphics[width =\textwidth]{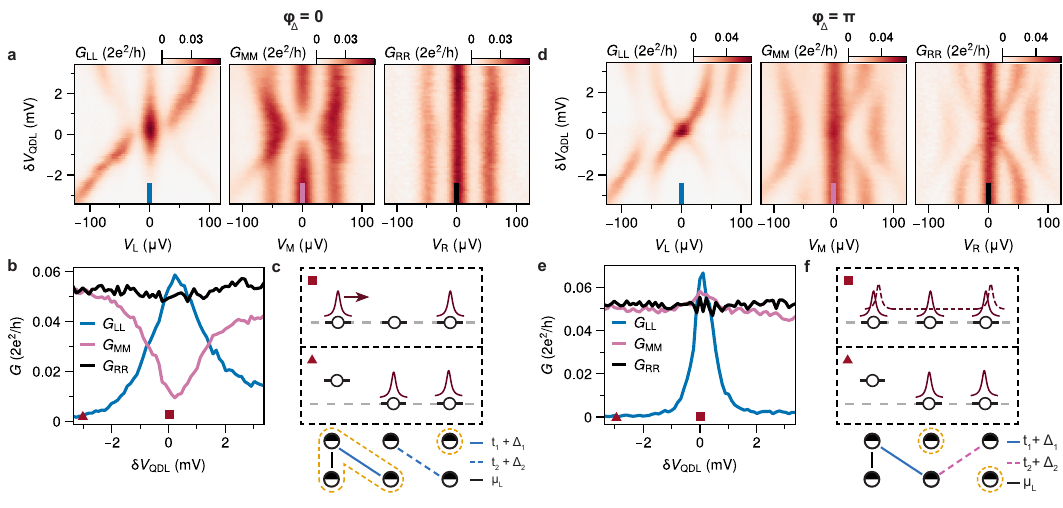}
    \caption{\textbf{Bulk protection of the MBS wavefunctions.} 
    \fl{a} Measured conductances with \var{B}{z} = \SI{21.4}{\micro\tesla} (corresponding to \phiD = 0 in Fig.~2) when sweeping \var{V}{QDL} around the charge degeneracy point. Measurements are performed with the same gate configuration as in Fig.~2.
    The ZBP in \var{G}{LL} gradually decreases in height, whereas a ZBP in \var{G}{MM} gradually appears.
    \fl{b} Zero-bias line-traces of conductance along indicated paths in (a).
    \fl{c} Visual representation of the measurements in (b), showing the MBS wavefunctions when \var{V}{QDL} is on resonance (square symbol) or off-resonance (triangle symbol). Bottom schematic shows the effect of introducing the coupling term $\mu_{\mathrm{L}}$, in the picture introduced in Fig.~2.
    \fl{d} Repetition of the measurement in (a) with \var{B}{z} = \SI{35.4}{\micro\tesla} (corresponding to \phiD~=~\uppi~). 
    The ZBP in \var{G}{LL} now splits from zero energy as \var{V}{QDL} is detuned. 
    The ZBPs in \var{G}{MM} and \var{G}{RR} remain, as the system effectively becomes a two-site chain as \var{V}{QDL} is detuned.
    \fl{e} Zero-bias line-traces of conductance along indicated paths in (d).
    \fl{f} Visual representation of the measurements in (e), detailed in text.
    The smooth control over \phiD~through \var{B}{z} allows for comparing these conductance spectra at any intermediate phase, shown in Fig.~S6.}
    \label{fig:Fig3}
\end{figure*}
\begin{figure}
\centering
    \includegraphics[width = 0.48\textwidth]{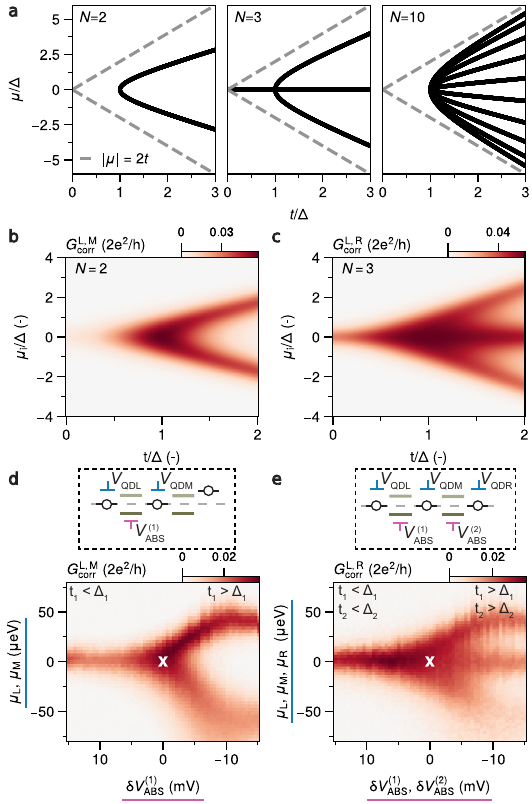}
    \caption{\textbf{The finite Kitaev chain's phase space.} \fl{a} Analytical zero-energy solutions to the N-site Kitaev chain with fixed $\Delta$, for different N (reproduced from \cite{Leumer2020}).
    The grey dashed line indicates the boundary of the topological phase formed at $N\to\infty$. 
    Numerically calculated conductance is shown for \fl{b} N~=~2 and \fl{c} N~=~3. 
    Here, \var{\Delta}~=~\SI{20}{\micro\electronvolt}.
    \fl{d} Measured correlated conductance when sweeping \ABS{(1)} against \var{V}{QDL} and \var{V}{QDM} (converted to \mui{L} and \mui{M} using QD leverarms).
    \fl{e} Measured correlated conductance when sweeping both \ABS{(1)} and \ABS{(2)} against \var{V}{QDL}, \var{V}{QDM} and \var{V}{QDR}.
    White 'x' marks the point in parameter space referred to as the sweet spot for the two and three site chain.
    In (d) and (e), x-axis are flipped to match numerical simulations. 
    The $\uparrow\uparrow\uparrow$ spin-configuration is used. 
    Measurements are reproduced in Fig.~S11.
    }
    \label{fig:Fig4}
\end{figure}

\section*{Results}
\subsection*{Tuning the two-site pairs}
Tuning a three-site Kitaev chain amounts to pairwise tuning of the two-site sweet spots (\var{t}{i}~=~\var{\Delta}{i})~\cite{DasSarma2012}.
These conditions can be inferred from zero-bias charge stability diagrams (CSDs) for the two pairs of QD resonances~\cite{Leijnse2012}, using the methods employed for two-site Kitaev chains~\cite{Dvir2023,Zatelli2023,tenHaaf2024}.
First, we obtain CSDs for the left and middle QDs, while keeping the right QD in Coulomb blockade (Fig.~1c).
The measurement is repeated as \ABS{(1)} is varied, to locate when the avoided crossing in the CSD changes direction.
A diagonal avoided crossing indicates $\Delta_1>t_1$ (Fig.~1c.i), whereas an antidiagonal avoided crossing results from $t_1>\Delta_1$ (Fig.~1c.iii). 
At an intermediate value of \ABS{(1)}, the desired $t_1=\Delta_1$ condition is satisfied, where the avoided crossing disappears (Fig.~1c.ii).
This procedure is repeated for the middle and right QDs, now varying \ABS{(2)}. 
This is seen in Fig.~1d, where \var{t}{2}~=~\var{\Delta}{2} is satisfied in panel 1d.ii.

It has been shown that robust, correlated ZBPs arise at these two-site sweet spots~\cite{Dvir2023, Zatelli2023, tenHaaf2024}.
These are manifestations of localised MBSs, in the two-site case referred to as ``Poor Man's Majoranas" as their stability is limited to single QD perturbations~\cite{Leijnse2012}.
We demonstrate this stability of the ZBPs for each pair of QDs.
In Fig.~1e we measure \var{G}{LL} as QDL is detuned, for the configuration in Fig.~1c.ii.
Similarly, \var{G}{RR} is measured for the configuration in Fig.~1d.ii upon varying \var{V}{QDR} (Fig.~1f).
In both cases we find robust zero energy states.
At the charge degeneracy point, the higher-energy excitations provide an estimate~\cite{Leijnse2012} for \var{t}{1}~=~\var{\Delta}{1}~$\approx$~\SI{20}{\micro\electronvolt} (Fig.~1e.ii) and \var{t}{2}~=~\var{\Delta}{2}~$\approx$~\SI{25}{\micro\electronvolt} (Fig.~1f.ii).
These measurements are in good agreement with the theoretical spectrum for well-polarised QDs~\cite{Tsintzis2022, Bozkurt2024, Liu2024Enhancing}, signifying that the ingredients for a full three-site Kitaev chain are present.

\subsection*{Phase control and the bulk excitation gap}
With \var{t}{} and \var{\Delta}{} balanced for each QD pair, the full three-site chain can be investigated.
Now the phase difference between the superconductors becomes a relevant parameter since it can affect the relative phase between the interdot couplings~\cite{DasSarma2012,Fulga2013, Bordin2024threesite}.
To investigate this, we set \ABS{(1)} and \ABS{(2)} to the sweet spot values obtained in Fig.~1 and set the QDs to their charge degeneracy points.
The conductance spectrum of each QD is then measured as a function of \var{B}{z} (Fig.~2a).
ZBPs are continuously observed on the outer QDs (\var{G}{LL} and \var{G}{RR}), while the middle QD (\var{G}{MM}) hosts higher energy excitations that move down to zero energy periodically.
This period in \var{B}{z} is equal to the period of \var{\phi}{SC} (\SI{28} {\micro\tesla}) measured for the bare superconducting junction.
We compare this behaviour to numerically calculated conductance spectra (Fig.~2b) of the three-site Kitaev chain as a function of \phiD, using the experimental coupling amplitudes estimated in Fig.~1.
A clear correspondence to the experimental result is obtained.
When the phase is 0 (modulo 2\uppi~), ZBPs are present in \var{G}{LL} and \var{G}{RR}, while an excitation gap is present in \var{G}{MM}.
As the phase is tuned towards $\pi$, finite-energy excitations lower in energy until the excitation gap is closed.

The observed behaviour can be understood when considering the Kitaev chain model in the Majorana basis (see Methods), visualised in Fig.~2c. 
Each QD site can be represented by two Majoranas.
For \phiD~=~0, an uncoupled Majorana arises only on the left and right sites, while the middle site is gapped. 
Experimental line-traces (Fig.~2d) at the corresponding magnetic field value are in agreement with this interpretation. 
A ZBP is present in \var{G}{LL} and \var{G}{RR} (panels i and iii), whereas \var{G}{MM} only shows excitations at higher bias (panel ii). 
This is a direct indication of the localisation of the MBS wavefunctions on the outer QDs.
In stark contrast, at a magnetic field that corresponds to \phiD~=~\uppi~, ZBPs are found on all three sites (Fig.~2e). 
Here, the coupling between neighbouring Majoranas is rearranged (see Fig.~2c), resulting in four uncoupled Majorana modes: one localised on each site, and an additional delocalized mode extending across all three sites.
Consequently, every site has a finite density of states at zero bias, leading to the observed conductances.

\subsection*{Shifting the Majorana wavefunctions}
The presence or absence of the excitation gap in the middle QD has direct consequences for the robustness of the ZBPs on the outer QDs.
In an ideal system, the energy of each isolated MBS cannot be lifted due to particle-hole symmetry.
Only an overlap between the two MBSs can achieve this, which is prevented by an excitation gap.
As shown above, the excitation gap is affected by \var{\phi}{\Delta}. 
The robustness of the ZBPs is therefore also expected to be phase-dependent. 
We compare the conductance spectra obtained at \var{B}{z} values corresponding to \var{\phi}{\Delta}~=~0 and \var{\phi}{\Delta}~=~\uppi~, where the excitation gap is either present or closed.

We first consider the spectra at \var{\phi}{\Delta}~=~0, upon detuning QDL (Fig.~3a).
The ZBP in \var{G}{LL} remains at zero bias, while a ZBP gradually appears in \var{G}{MM}. 
The conductances at zero bias are expected to be proportional to the density of the Majorana wavefunction at each site (see Fig.~S8). 
Fig.~3b shows extracted line-traces at zero bias. 
As QDL is detuned, \var{G}{LL} gradually reduces, while \var{G}{MM} simultaneously increases. 
This happens because the weight of the Majorana wavefunction shifts from the left to the middle QD (illustrated in Fig.~3c)~\cite{pandey2025}. 
Finally, we see that the spectrum at QDR is hardly affected, a consequence of the isolation of the MBS on this site.

The spectra at \var{\phi}{\Delta}~=~\uppi~, where the gap in QDM is closed, have a markedly different behaviour~(Fig.~3d). 
The ZBP in \var{G}{LL} now splits immediately in energy as QDL is detuned.
This is indeed expected, since MBSs in the system now directly overlap, and are thus no longer robust to detuning. 
Line-traces at zero bias (Fig.~3e) show two important features. 
Firstly, \var{G}{LL} now drops faster compared to Fig.~3b, due to the lack of protection. 
Secondly, the ZBPs in both \var{G}{MM} and \var{G}{RR} are now unaffected by the detuning of the left QD. 
This can again be understood in terms of the MBS wavefunctions (illustrated in Fig.~3f), whereby detuning the left QD now fuses the localised MBS on QDL with the delocalised MBS extending across the chain~\cite{pandey2023fusion,Pandey2023_yshape, Liu2024scaling}. 
These observations, and conductance spectra obtained for additional combinations of QD detunings, again show a striking agreement with numerical simulations (Fig.~S7, S9).

\subsection*{ZBPs outside the sweet spot}
In previous sections the system was characterised at fine-tuned sweet spots where ZBPs appear on the outer QDs. 
It is expected that ZBPs should also be present away from these sweet spots. 
This follows from a key property of the Kitaev model, whereby the region in parameter space that hosts MBSs grows as the number of sites $N$ is increased.
For finite chains, MBSs are fully localised at the edges only at the \var{t}{i} = \var{\Delta}{i} sweet spot~\cite{Leumer2020}. 
However, ZBPs arising from delocalized MBSs are expected away from this point.
To study this, it is instructive to extract zero-energy solutions of the Kitaev chain as a function of $\mu$ and $t$, for fixed $\Delta$.
An exact analytical expression for the zero-energy solutions exists for any $N$~\cite{Leumer2020}, shown in Fig.~4a. 
For $N=2$, the solutions lie on a single hyperbola, with a vertex at $t=\Delta$,~$\mu=0$ (i.e. the two-site sweet spot). 
For each odd \var{N}{}, $\mu=0$ is always a solution~\cite{Ezawa2024_even_odd} (see \var{N}{}~=~3).
As \var{N}{} increases, the number of solutions increases and gradually fills the region where $\lvert \mu \rvert < 2t$.
As $N\to \infty$, this region is filled with states exponentially close to zero energy, ultimately creating a topological phase within which the MBSs are always localised at the edges.
The zero-energy solutions can be studied by measuring zero-bias conductances on each QD while varying both interdot couplings and the QD electrochemical potential energies.
Fig.~4b and Fig.~4c show numerical simulations, for \var{N}{}~=~2 and \var{N}{}~=~3 respectively.

Experimentally this parameter space is only indirectly accessible, as \var{t}{i} and \var{\Delta}{i} are not controlled independently.
However, starting with the hybrid gates at their sweet spot value (denoted $\delta$\ABS{(i)}~=~0) and sweeping both simultaneously, the system can be smoothly tuned between \var{t}{i}~$<$~\var{\Delta}{i} and \var{t}{i}~$>$~\var{\Delta}{i} regimes (full procedure in Fig.~S10).
Fig.~4d shows this for the left and middle QD pair. 
Here, the conductance features meeting at $\delta$\ABS{(i)}~=~0 agree qualitatively with the behaviour expected from Fig.~4b.
Instead of extending linearly as \var{t}{1}~$>$~\var{\Delta}{1}, the feature saturates in \mui{i}, since $t_{1}$ does not increase independently of \var{\Delta}{1} in these experiments. 
The finite conductance `tail' at  \mui{i}~=~0 here is a consequence of both \var{t}{} and \var{\Delta}{} decreasing in amplitude (see Fig. S10), which is not captured by the theory model.

Next, the measurement is performed for the full chain, sweeping simultaneously both hybrid gates against detuning all three QDs (Fig.~4e).
The appearance of the additional conductance feature along \var{\mu}{i}~=0 marks the first step of the growth into a topological phase in this discrete system and agrees well with the Kitaev chain model in a large parameter space. 
Finally, these measurements allow us to qualitatively visualise the increase in stability of the MBSs as the chain is extended~\cite{Bordin2024threesite}. 
We find that conductance decays more slowly for the three-site chain compared to the two-site chain as one moves away from the sweet spot (marked by white `x') in any direction.

\section*{Conclusions}
In summary, we have realised a three-site Kitaev chain by coupling three QDs via two semiconductor-superconductor regions. 
We find that the phase difference between the superconductors controls the relative phase between the interdot couplings, and thereby the energy spectrum of the system.
This finding is particularly relevant for the study of zero energy modes in longer chains~\cite{luethi2024fatepoormansmajoranas, svensson2024quantumdotquality}, where control over these phases becomes crucial~\cite{DasSarma2012,Liu2024scaling}.
By appropriately tuning the phase, we show that the appearance of ZBPs on the outer QDs is accompanied by an excitation gap in the middle QD, evidence of the zero-energy modes being localised at the edges.
We further demonstrate that detuning of the QDs in the chain allows one to smoothly transfer the weight of the Majorana wavefunction between different sites.  
This kind of spatial manipulation of MBSs is a crucial requirement in gate-based braiding proposals for realising a Majorana qubit~\cite{Alicea2011_braiding, Palyi2024_braiding_based_control}.
Overall, our experiments agree well with the predictions of the Kitaev model in a wide parameter space and are an important step towards studies that require a reliable way to produce robust, localised Majorana bound states~\cite{tsintzis2023,Liu2023b_fusion,Pan2024Rabi}.

\section*{Acknowledgements}
We thank O.W.B.~Benningshof and J.D.~Mensingh for technical assistance with the cryogenic electronics.
We thank L.~Vandersypen, R.~Aguado, R.~Seoane, M.~Leijnse, G.~Wang, A.M.~Bozkurt, N.~van Loo, G.P.~Mazur, F.~Zatelli and L.~Kouwenhoven for providing valuable input on the manuscript. 
The research at Delft was supported by the Dutch National Science Foundation (NWO),  Microsoft Corporation Station Q and a grant from Top consortium for Knowledge and Innovation program (TKI). 
S.G. and M.W acknowledge financial support from the Horizon Europe Framework Program of the European Commission through the European Innovation Council Pathfinder grant no. 101115315 (QuKiT).

\section*{Author contributions}
Q.W. fabricated the device, with input from S.L.D.tH and I.K. 
Y.Z. and I.K. fabricated the resonator circuits.
C.G.P. and I.K. designed the measurement set-up.
Measurements were performed by S.L.D.tH,  V.S.,  Y.Z. and A.B..
C.-X.L. and M.W. provided the theoretical analysis, with C.-X.L. carrying out the analytical calculations and the numerical simulations.
MBE growth of the semiconductor heterostructures and the characterisation of the materials were performed by D.X. and C.T. under the supervision of M.J.M.
The manuscript was written by S.L.D.tH. and S.G., with input from all co-authors. 
S.G. supervised the experimental work in Delft.

\section*{Data availability}
All raw data obtained in relation to this manuscript and the scripts to produce the figures are available on Zenodo~\cite{tenHaafDataRepo}. 

\bibliography{ref} 

\end{document}


\title{Supplementary Material: Observation of edge and bulk states in a three-site Kitaev chain}

\maketitle

\tableofcontents
\clearpage
\section{Methods - Experiments}

\subsection{Fabrication and yield}
All devices were fabricated using the techniques described in detail in \cite{Moehle2021} and \cite{WangQThesis}. 
Aluminium loop structures are defined in an InSbAs-Al chip by wet etching, followed by the deposition of three ohmic Ti/Pd contacts. 
After deposition of \SI{20}{nm} AlOx via \SI{40}{\celsius} atomic layer deposition (ALD), three large Ti/Pd depletion gates are evaporated: one large top depletion gate and two bottom depletion gates each extending halfway.
The bottom consists of two gates in order to independently form the left and right halves of the channel, separated by a thin channel where the middle lead is placed.
This simplifies placing an additional gate in front of the middle ohmic contact, in order to have a well defined tunneling barrier for the middle probe.  
Additionally, it simplifies finding a suitable active region where the 2DEG is depleted below the gates, while a narrow conductance channel remains in between.
The channel width is designed to be \SI{200}{\nano\meter}.
Following a second ALD layer (\SI{20}{\nano\meter} AlOx), a first layer of six Ti/Pd finger gates is evaporated.
These are used for controlling the electrochemical potential energies of the QDs and hybrid regions and for defining the tunneling barrier for the middle contact. 
A third ALD layer (\SI{20}{\nano\meter} AlOx) is deposited, followed by evaporation of the remaining six Ti/Pd finger gates that define the three QDs.
For RF-measurements, superconducting LC-resonator circuits are fabricated on a separate chip with a silicon substrate, by etching NbTiN.
To apply DC voltages, bias tees are created by depositing \SI{20}{\nano\meter} Cr structures with resistances of $\approx$~\SI{5}{\kilo\ohm}.
\newline

For the investigation presented in this manuscript, in total sixteen devices were fabricated, with small variations in device dimensions.
Of these, twelve had no visible defects under optical/SEM inspection of the finished devices.
We faced some challenges with simultaneously connecting all 15 gates to the printed circuit board and connecting all 3 ohmic contacts via the resonator chip.
In total six devices were bonded and cooled down, of which in five instances fridge wiring and bonding issues caused gate or ohmic connections to be absent once reaching milli-Kelvin temperatures.
Once these issues were resolved, the final device was used for obtaining the measurements demonstrated in this manuscript. 
This device was cooled down twice over the course of five months, in which we were able to reproduce our findings when re-starting from a `reset' device.
The main text highlights results within a single cooldown, data from the other sets of experiments is included in this supplementary.

\subsection{DC transport measurements}
Measurements are performed in a dilution refrigerator with a base temperature of \SI{20}{\milli\kelvin}.
Transport measurements presented in the main text are performed in AC and DC using a four-terminal set-up (three ohmic contacts plus two aluminium strips connected in a loop).
The aluminium strips induce a gap of $\approx$~\SI{220}{\micro\volt} (Fig.~\ref{supp:CharactABss}) and are kept electrically grounded. 
Each ohmic lead is connected to a current meter and biased through a digital-to-analogue converter and both DC and AC voltages can be applied.
Offsets of the applied voltage-bias on each lead are corrected via independently measuring the Coulomb peaks in the QDs and looking at the change in sign of the current.
The voltage outputs of the current meters are recorded with three digital multimeters and three lock-in amplifiers.  
When applying a DC voltage to one lead (e.g., \var{V}{L}),  the other leads (i.e., \var{V}{M} and \var{V}{R}) are kept grounded.
AC excitations are applied with amplitudes around \SI{5}{\micro \volt} RMS and a frequency of \SI{23}{\hertz}.
In this way, a full conductance matrix $G_{ij}=\frac{dI_{i}}{dV_{j}}$ is obtained by measuring the response of \var{I}{L}, \var{I}{M} and \var{I}{R}, to \var{V}{L}, \var{V}{M} and \var{V}{R}.
Three separate measurements are required, as only a single lead is biased at a time. 
Small offsets in measured conductances arise using the lock-in amplifiers, due to capacitances to ground within the electronics.
These offsets are calibrated using Coulomb blockaded measurements and corrected. 
It should be noted that voltage-divider effects arise when applying biases in a four-terminal set-up.
For three-terminal set-ups, the measured conductances and applied biases are generally corrected~\cite{martinez2021}. 
For a four-terminal set-up, however, such a calculation gets cumbersome.
Here, we focus on low tunneling regimes (G $\ll$ 2e$^2$/h) where the device resistance is large compared to the resistances of the fridge lines and the current meters, such that the multi-terminal effect is small.
Nevertheless, this should be kept in mind when e.g. interpreting the non-local conductances in Fig.~\ref{supp:fullMatrix}. 

Magnetic fields are applied using a 3D vector magnet. 
The field perpendicular to the superconducting loop (\var{B}{z}) is generated using a high-resolution current source, giving a \var{B}{z} resolution below \SI{0.1}{\micro\tesla} (providing sufficient resolution for the flux period of \SI{28}{\micro\tesla}). 
A small (but significant) hysteresis on the order of \SI{5}{\micro\tesla} is observed when sweeping \var{B}{z} in opposite directions.
This is counteracted by setting \var{B}{z} first to \SI{-100}{\micro\tesla} and then sweeping this field back in the positive direction, such that consecutive experiments where \var{B}{z} is varied are consistent.
To spin-polarise the QDs, a magnetic field of \SI{200}{\milli\tesla} is applied parallel to the channel (\var{B}{x}).
Due to an imperfection in the alignment, this introduces a small \var{B}{z} component as well, on the order of 80 flux quanta. 
It was not possible to accurately correct this offset for this work, and so we do not determine the \var{B}{x} value that corresponds to precisely $0$ flux through the loop.
 
\subsection{RF-reflectometry measurements}
The experiments require the tuning of 15 gate voltages in order to create the three-site chain, which results in a large parameter space of gate voltages.
To speed up the tuning process we employ radio-frequency (RF) lead reflectometry~\cite{Vigneau2023} in addition to the DC conductance measurements. 
Each Ohmic contact is connected to an inductor, designed with varying inductances $L_{\mathrm{L, M, R}} = 0.2, 0.5, 1.5$~\SI{}{\micro \henry}, that together with a parasitic capacitance to ground via bond-wires result in resonators with frequencies of $f_{\mathrm{L, M, R}} = 723, 505, 248$~\SI{}{\mega\hertz}.
A complete circuit diagram including the fridge wiring and filters is provided in~\cite{Kulesh2024}.
Using a directional coupler, we obtain the reflected signal of each lead.
We denote with $S_{21}^{\mathrm{L}}$,  $S_{21}^{\mathrm{M}}$ and  $S_{21}^{\mathrm{R}}$ the normalised reflected signals of the left, middle and right lead respectively, which correspond roughly linearly to conductance.
All three signals can be measured simultaneously through multiplexing~\cite{Hornibrook2014}, using the circuit shown in Fig.~\ref{supp:CharactABss}. 
In combination with saw-tooth pulses on the QD plunger gates, generated by arbitrary waveform generators, this allows for scanning the parameter space many times faster than through DC measurements.
For clarity, throughout the supplementary information the data obtained with RF reflectometry is displayed in a different colormap from the DC conductance data.

\subsection{Device tune-up}
Here, we detail how the device was tuned to reach the point of obtaining the measurements presented in the main text.
To start, a conductance channel has to be isolated, by tuning the three large depletion gates. 
A regime is obtained where the 2DEG is fully depleted below these gates, but a finite current remains in the channel.
Next, we perform tunnel spectroscopy of the hybrid regions to verify the presence of gate-tunable sub-gap states (Fig.~\ref{supp:CharactABss}). 
Then, the remaining gates are activated to form the three QDs.
We find a regime in which the QDs are stable and contain orbitals with a large level spacing, such that sufficient Zeeman energies can be induced through the external magnetic field without mixing states in neighbouring orbitals.
The InSbAs hetero-structure investigated in this manuscript does not have a capping layer, such that the confinement layer is very close to the surface~\cite{Moehle2021}. Thus the QDs can couple easily to disorder in the dielectric, resulting in regimes in parameter space that are unstable. This constitutes a challenge in tuning up all 3 QDs into stable regimes~(Fig.~\ref{supp:S3stability}). Once such a stable regime is found, we tune the interdot couplings by studying charge stability diagrams of the QDs and the ABSs in the hybrid regions, following the protocols in~\cite{Liu2024Enhancing,Zatelli2023}. 
An important requirement is that the interdot coupling amplitudes should exceed the line-width of the conductance measurements, in order to clearly resolve the full density of states.
However, if the middle QD is too strongly coupled to both hybrids, its chemical potential energy becomes dependent on the superconducting phase~\cite{Aguado2024}. 
In this case, the interdot coupling amplitudes (\var{t}{}, \var{\Delta}{}) themselves become phase dependent and additional modulations in the conductance spectra are observed (see Fig.~\ref{fig:bulkgapfit}e).
We limited the coupling between the ABSs and the middle QD in order to stay away from this regime (as verified by Fig.~\ref{supp:PMMspectra}d,h), in order to compare the conductance spectra as a function of \var{B}{z} directly to the Kitaev chain model in Fig.~2. 
For the conductance spectra measurements shown in Fig.~2 and Fig.~3, it is crucial that the QDs remain on resonance throughout the measurements. 
The center of each resonance was determined by measuring independently the Coulomb resonances for each QD in RF-reflectometry and fitting a Lorentzian line-shape. 
This was done directly before and after each measurement, to verify that no gate jump or drift occurred.
Additionally, CSDs for the left and right QD pairs were obtained before and after each set of measurements, to ensure that the interdot couplings had not changed.
In case any gate-jumps occurred, the measurements were repeated.

\section{Methods - Theory}
\subsection{Kitaev model and numerical calculations of conductance}
The Hamiltonian of a three-site Kitaev chain is given by:
\begin{align}
H = \mu_1 n_1 + \mu_2 n_2 + \mu_3 n_3 +  (t_1 c\dg_2 c_1 + t_2 c\dg_3 c_2 + \Delta_1 c_2 c_1 + \Delta_2 e^{i\phi_{\Delta}} c_3 c_2 + h.c.),
\end{align}
where $c_i$ is the annihilation operator for the fermion on the $i$-th site, $n_i=c\dg_i c_i$ is the number operator, $\mu_i$ is the onsite energy, $t_i$ and $\Delta_i$ are the normal and superconducting tunneling amplitudes between neighbouring sites, and $\phi_{\Delta}$ is the phase difference between the two superconducting leads which can be controlled by the magnetic flux threading through the loop.
The corresponding Bogoliubov-de-Gennes Hamiltonian is 
\begin{align}
h_{K3} = 
\begin{pmatrix}
\mu_1  &  t_1  &  0 &  0   & \Delta_1 & 0\\
t_1  &  \mu_2  &  t_2 &  -\Delta_1   & 0 & \Delta_2 e^{-i\phi_{\Delta}}\\
0  &  t_2  &  \mu_3 &  0   & -\Delta_2 e^{-i\phi_{\Delta}} & 0\\
0  &  -\Delta_1  &  0 &  -\mu_1   & -t_1 & 0\\
\Delta_1  &  0  &  -\Delta_2 e^{i\phi_{\Delta}} &  -t_1   & -\mu_2 & -t_2\\
0  &  \Delta_2 e^{i\phi_{\Delta}}  &  0 &  0   & -t_2 & -\mu_3
\end{pmatrix}
\end{align}
under the basis of $(c\dg_1, c\dg_2, c\dg_3, c_1, c_2, c_3)$.
To calculate the differential conductance through the system, we use the scattering matrix method, where the $S$ matrix can be obtained from the Weidenmuller formula as below
\begin{align}
S(\omega)
=\bpm
S^{ee} & S^{eh}  \\
S^{he} & S^{hh} 
\epm
 = \hat{1} - i W\dg (\omega - h_{K3} + \frac{i}{2}WW\dg)^{-1}W,
\end{align}
where $W=\text{diag}(\sqrt{\Gamma_1}, \sqrt{\Gamma_2}, \ldots, -\sqrt{\Gamma_1}, -\sqrt{\Gamma_2}, \ldots)$ is the matrix describing the dot-lead couplings, with $\Gamma_i$ being the coupling strength between dot-$i$ and lead-$i$.
The zero-temperature conductance is thus
\begin{align}
G^{(0)}_{\alpha \beta}(\omega) = \delta_{\alpha \beta} - |S^{ee}_{\alpha \beta}(\omega)|^2 + |S^{he}_{\alpha \beta}(\omega)|^2
\end{align}
in unit of $e^2/h$.
The finite-temperature conductance is obtained by a convolution between the zero-temperature conductance and the derivative of the Fermi distribution
\begin{align}
G^{T}_{\alpha \beta}(\omega) = \int^{+\infty}_{-\infty} dE \frac{G^{(0)}_{\alpha \beta}(E)}{ 4k_BT \cosh^2[(E-\omega)/2k_BT] }.
\end{align}
In the numerical calculations shown in the current work, we use $t_1=\Delta_1=$~\SI{20}{\micro\electronvolt}, $t_2=\Delta_2=$~\SI{25}{\micro\electronvolt}, which are extracted from experimental data in Fig.~1.
In addition, we set $\Gamma_1=\Gamma_2=\Gamma_3=$~\SI{0.7}{\micro\electronvolt}, which does not affect the simulations qualitatively, but were selected to give the same order of magnitude of conductance. 
Lastly, we set $T$~=~\SI{50}{\milli\kelvin}.

\subsection{The Kitaev chain in the Majorana basis}
To guide interpretation of the measurements, it is instructive to consider the three-site Kitaev chain Hamiltonian in terms of Majorana operators. 
This is done by introducing two Majorana operators for each fermionic site:
\begin{align}
 c_n = ( \gamma_{na} + i \gamma_{nb} ) / \sqrt{2}, \quad c\dg_n = ( \gamma_{na} - i \gamma_{nb} ) / \sqrt{2}, \quad \{ \gamma_{ma}, \gamma_{nb} \} = \delta_{mn} \delta_{ab}.
\end{align}
As such, the Hamiltonian can be written as
\begin{align}
H &= i \mu_1 \gamma_{1a} \gamma_{1b} + i \mu_2 \gamma_{2a} \gamma_{2b} + i \mu_3 \gamma_{3a} \gamma_{3b} + i (t_1 + \Delta_1) \gamma_{2a} \gamma_{1b} + i (-t_1 + \Delta_1) \gamma_{2b}\gamma_{1a}  \nn
& + it_2 \gamma_{3a} \gamma_{2b} - it_2 \gamma_{3b} \gamma_{2a}
+i \Delta_2 \cos\phi (\gamma_{3a} \gamma_{2b} + \gamma_{3b} \gamma_{2a})
+i \Delta_2 \sin \phi (\gamma_{3a} \gamma_{2a} - \gamma_{3b} \gamma_{2b}).
\end{align}
Therefore, at the sweet spot of a three-site Kitaev chain, i.e., $\mu_n=0, t_1=\Delta_1, t_2=\Delta_2, \phi=0$, we have
\begin{align}
H(\phi=0) =  i2t_1 \gamma_{2a} \gamma_{1b} + i2t_2 \gamma_{3a} \gamma_{2b}.
\label{eq:Ham_0}
\end{align}
On the other hand, at $\mu_n=0, t_1=\Delta_1, t_2=\Delta_2, \phi=\pi$, the Hamiltonian becomes 
\begin{align}
H(\phi=\pi) =  i2t_1 \gamma_{2a} \gamma_{1b} - i2t_2 \gamma_{3b} \gamma_{2a}.
\end{align}
A bilinear term of Majorana operators indicates a coupling between them. This is what is visualised through the schematics used in the main text (i.e. Fig.~2c).

\subsection{Majorana zero modes for a Kitaev chain at $\pi=0$ and $\phi=\pi$}

We now calculate the wavefunctions of Majorana zero modes of the Kitaev chain.
To distinguish it from the Majorana basis using $\gamma$'s, we use $\chi_n$ to denote the Majorana zero modes. The definition of a Majorana zero-energy quasiparticle is:
\begin{align}
[H, \chi_n] = 0, \quad \chi\dg_n=\chi_n.
\end{align}
For the Kitaev chain with $\phi=0$ the solutions are easily obtained: since $\gamma_{1a}$ and $\gamma_{3b}$ do not appear in the Hamiltonian in Eq.~\eqref{eq:Ham_0}, the wavefunctions of the two Majorana zero modes are:
\begin{align}
\chi_1 = \gamma_{1a}, \quad \chi_2 = \gamma_{3b},
\end{align}
i.e., one is completely localised at the left QD and the other at the right QD, separated in space by the middle QD.
On the other hand, when $\phi=\pi$, there exist four Majorana zero modes, of which the first three are again easy to find as shown below:
\begin{align}
 \chi_1 = \gamma_{1a}, \quad \chi_2 = \gamma_{2b}, \quad \chi_3 = \gamma_{3a}.
\end{align}
They are localised at the left, middle and right QD respectively.
The ansatz of the fourth Majorana zero mode is $\chi_4 = A \gamma_{1b} + B\gamma_{2a}  + C \gamma_{3b}$, which yields:
\begin{align}
\chi_4 = \frac{t_2}{\sqrt{t^2_1 + t^2_2}} \gamma_{1b}  -\frac{t_1}{\sqrt{t^2_1 + t^2_2}}  \gamma_{3b}.
\end{align}
That is, $\chi_4$ is delocalised at the left and right QDs (no wavefunction weight in the middle dot), with the relative weights determined by the gap of the left and right pairs.
One can note that once the left or right QD is detuned from resonance, i.e., a finite $\mu_1$ or $\mu_3$, $\chi_4$ would couple with $\chi_1$ or $\chi_3$, giving an energy splitting. 
This explains the splitting of the zero-bias conductance peak observed in the $\pi$-phase Kitaev chain studied in the main text (Fig.~3d).

\subsection{The Majorana density}
The main text focuses on conductance spectra of the three-site Kitaev chain at $\phi_{\Delta}=0$, when the left QD is detuned. 
Here we show that an isolated zero-bias conductance peak reveals the Majorana wavefunction profiles.
Assuming equal dot-lead coupling strengths $\Gamma_i=\Gamma$ and an isolated zero-energy state of $h_{K3}$, the $S$ matrix can be simplified to:
\begin{align}
S(\omega) \approx \hat{1} - i \frac{\Gamma}{\omega + i \frac{\Gamma}{2}}
\bpm
u_1 u^*_1 + v^*_1 v_1, \ldots \\
u_2 u^*_1 + v^*_2 v_1, \dots \\
\vdots \\
-(v_1 u^*_1 + u^*_1 v_1), \ldots \\
-(v_2 u^*_1 + u^*_2 v_1), \dots \\
\vdots \\
\epm
\end{align}
where $u_i$ and $v_i$ are the electron and hole components of the zero-energy quasiparticle state on site-$i$.
By definition $ u^* = ( \xi_A + i\xi_B )/\sqrt{2}, v = ( \xi_A - i\xi_B )/\sqrt{2}$, where $\xi_A$ and $\xi_B$ are Majorana wavefunctions.
We thus have $ u u^* + v^* v = |\xi_A|^2 + |\xi_B|^2,  -2u^*v = -(\xi^2_A + \xi^2_B).$
Furthermore, $h_{K3}$ is real at $\phi=0$, giving real $u$ and $v$.
Thereby $\xi_A$ is purely real and $\xi_B$ is purely imaginary, giving $ u u^* + v^* v = \rho_A + \rho_B,  -2u^*v = -(\rho_A - \rho_B)$,
where $\rho_{A/B} = |\xi_{A/B}|^2$ are the local Majorana densities.
As a result, the local conductance in the zero-temperature limit is:
\begin{align}
&G^{(0)}_{ii}(\omega) = \frac{ (\Gamma/2)^2 }{\omega^2 + (\Gamma/2)^2} \cdot 4(\rho_{Ai} + \rho_{Bi} - 4\rho_{Ai}\rho_{Bi}).
\end{align}
Here, the profile of the zero-bias conductance peak has a Lorentzian shape with its broadening width being fixed by the dot-lead coupling strength.
Although finite-temperature effects will change this prefactor, a key finding here is that the local zero-bias conductance of a multi-terminal junction is proportional to the Majorana densities.

For a three-site Kitaev chain at its sweet spot, as considered in the current work, the zero-energy eigenfunction is $\psi = \frac{1}{2} (\frac{1}{\sqrt{1+4a^2}}, \frac{-2a}{\sqrt{1+4a^2}}, 1, \frac{-1}{\sqrt{1+4a^2}},\frac{2a}{\sqrt{1+4a^2}}, 1 )^T$, giving $\rho_A=(0,0,1/2)$ and $\rho_B=\frac{1}{2}(\frac{1}{1+4a^2}, \frac{4a^2}{1+4a^2},0)$ where $a=\mu_1/4t$ is the dot detuning with respect to gap.
Therefore, the profiles of local zero-bias conductances as a function of dot detuning are:
\begin{align}
& G_{11}(V=0) \propto 2\rho_B = \frac{1}{1+4a^2}, \nn
& G_{22}(V=0) \propto 2\rho_B = \frac{4a^2}{1+4a^2}, \nn
& G_{33}(V=0) \propto 2\rho_A = 1.
\label{eq:detuning}
\end{align}
Thus, the local conductance profiles as a function of dot detuning indicate the moving of Majorana zero modes in a three-site chain.
We compare these findings to the experimental data in Fig.~\ref{supp:detuningvsanalytics}.

\subsection{Line-shape of conductance traces}
When probing the excitation gap in the middle QD at the three-site sweet spot (Fig.~2), a finite in-gap conductance remains. 
In part, this can be expected due to thermal-broadening effects on the higher energy excitations. 
We address here the expected line-shapes for the conductances at a three-site sweet spot.
The conductance measurements are performed in the limit of low tunneling between the leads and the QDs, such that we can expect the QD-lead coupling $\mathit{\Gamma} \ll \mathrm{k}_{\mathrm{B}}T$.
In this limit, the line-shape of a single QD resonance as a function of applied $V_{\mathrm{bias}}$ is given by~\cite{Beenakker1991}:
\begin{align}
G(V_{bias}) \propto A\cosh{\left(\frac{V_{\mathrm{bias}}}{2k_{\mathrm{B}}T}\right)}^{-2}
\end{align}
In conductance measurements of a three-site chain sweet-spot, the density of states depends on the site that is probed.
For the left QD, three conductance peaks are expected: one at zero bias, and two at $\pm$~$2t_1$.
We thus model the conductance \var{G}{LL} as a function of \var{V}{L} to be the sum of 3 contributions:
\begin{align}
G_{\mathrm{LL}}(V_{\mathrm{L}}) = A_1\cosh{\left(\frac{V_{\mathrm{L}}-2t_1}{\gamma}\right)}^{-2} 
+
A_2\cosh{\left(\frac{V_{\mathrm{L}}}{\gamma}\right)}^{-2} 
+
A_3\cosh{\left(\frac{V_{\mathrm{L}}+2t_1}{\gamma}\right)}^{-2} 
\label{eq:Gshape_GLL}
\end{align}
Where we replace the factor of $2k_{\mathrm{B}}T$ with a general broadening factor $\gamma$.
Similarly, the right QD has three conductance peaks: one at zero bias, and two at $\pm$~$2t_2$. 
We model the conductance \var{G}{RR} as a function of \var{V}{R} to be:
\begin{align}
G_{\mathrm{RR}}(V_{\mathrm{R}}) = B_1\cosh{\left(\frac{V_{\mathrm{R}}-2t_2}{\gamma}\right)}^{-2} 
+
B_2\cosh{\left(\frac{V_{\mathrm{R}}}{\gamma}\right)}^{-2} 
+
B_3\cosh{\left(\frac{V_{\mathrm{R}}+2t_2}{\gamma}\right)}^{-2} 
\label{eq:Gshape_GRR}
\end{align}
From these fits, the values of $t_{1}$ and $t_{2}$ are determined. 
The conductance of the middle QD is then expected to have 4 peaks: at $\pm t_1$ and at $\pm t_2$, giving the sum of the following four contributions:
\begin{align}
G_{\mathrm{MM}}(V_{\mathrm{M}}) = C_1\cosh{\left(\frac{V_{\mathrm{M}}-2t_1}{\gamma}\right)}^{-2} 
+
C_2\cosh{\left(\frac{V_{\mathrm{M}}+2t_1}{\gamma}\right)}^{-2} 
+
C_3\cosh{\left(\frac{V_{\mathrm{M}}-2t_2}{\gamma}\right)}^{-2} 
+
C_4\cosh{\left(\frac{V_{\mathrm{M}}+2t_2}{\gamma}\right)}^{-2} 
\label{eq:Gshape_GMM}
\end{align}
These analytical expressions are in close agreement with the conductance traces obtained from the numerical simulations, in the parameter regime described above. We compare the expressions to experimental data in Fig.~\ref{fig:bulkgapfit}.

\setcounter{figure}{0}
\renewcommand{\thefigure}{S\arabic{figure}}
\newpage
\section{Supplementary Figures S1 to S11}

\begin{figure}[h!]
\centering
\includegraphics[width=\textwidth]{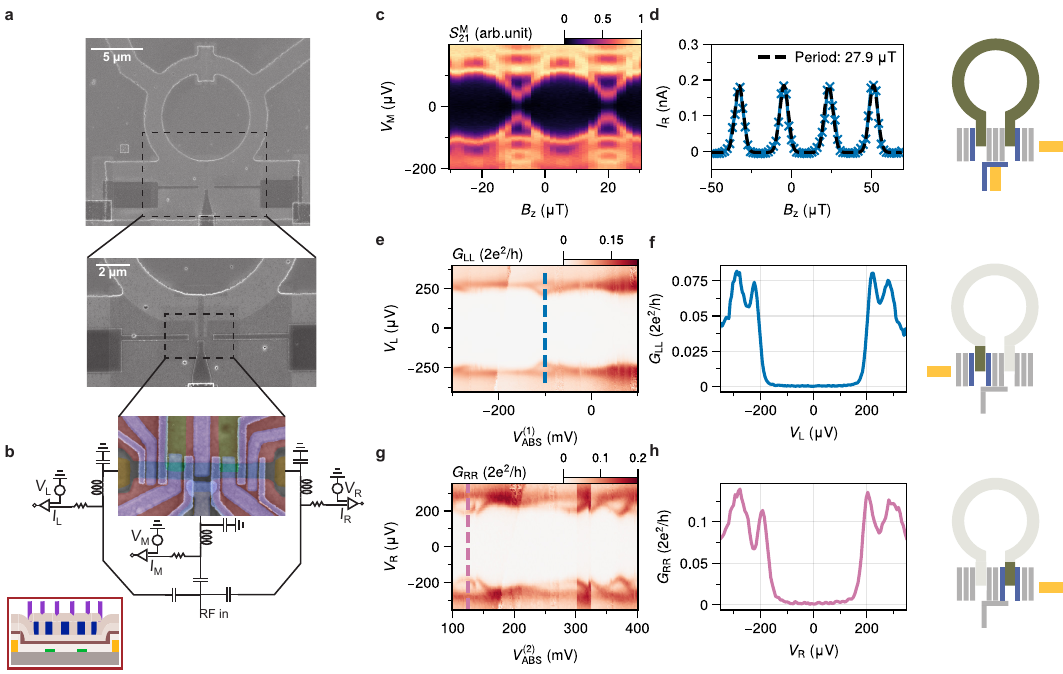}
\caption{\textbf{Device description  and characterisation of hybrid sections.} \fl{a} Zoomed out SEMs of copies of the measured device obtained after deposition of Ohmic contacts, showing the full structure of the superconducting loop. 
\fl{b}~Close-up SEM of the finished device, including the full circuit diagram. 
Bottom left in-set shows a cross-sectional schematic of the device along the channel, adapted from~\cite{WangQThesis}, to visualise the order of the three gate layers. 
Resonators are formed by inductors in combination with a parasitic capacitances to ground, which allows for fast radio-frequency (RF) measurements, used in this work for tuning and characterisation of the system. 
Voltage sources and current meters are attached to each lead via $\approx$~\SI{5}{\kilo\ohm} resistors, acting as bias tees, used to obtain the conductance measurements in the main text. 
Sub-figures (c-h) show characterisations of the three possible hybrid configurations, after a 1-D channel is formed with the large depletion gates. 
Schematics on the right display the activated tunneling gates and relevant Ohmic contacts for each row.
\fl{c} RF-spectroscopy of the Josephson junction formed by the two superconducting fingers at the ends of the loop, as a function of the magnetic field $\mathrm{B}_{\mathrm{z}}$ perpendicular to the loop. 
\fl{d} Measured current \var{I}{R} with \var{V}{R}=\SI{100}{\micro\volt}, in a wider range of \var{B}{z}.
Fitting the oscillations with a periodic Gaussian function provides an estimate for the flux periodicity (\SI{28}{\micro\tesla}). 
\fl{e} Tunneling spectroscopy of the left SC finger in isolation, as a function of 
\ABS{(1)} applied to the gate covering the left hybrid region. 
\fl{f} Line-trace from (e) at \ABS{(1)}=\SI{-100}{\milli\volt}, to show the presence of a sub-gap state in the left hybrid.
\fl{g} Tunneling spectroscopy of the right SC finger in isolation, as a function of the gate covering the right hybrid region (\ABS{(2)}). 
\fl{h} Line-trace from (g) at \ABS{(2)}=\SI{125}{\milli\volt} to show the presence of a sub-gap state in the right hybrid.
}
\label{supp:CharactABss}
\end{figure}

\begin{figure}[t!]
\centering
\includegraphics[width=\textwidth]{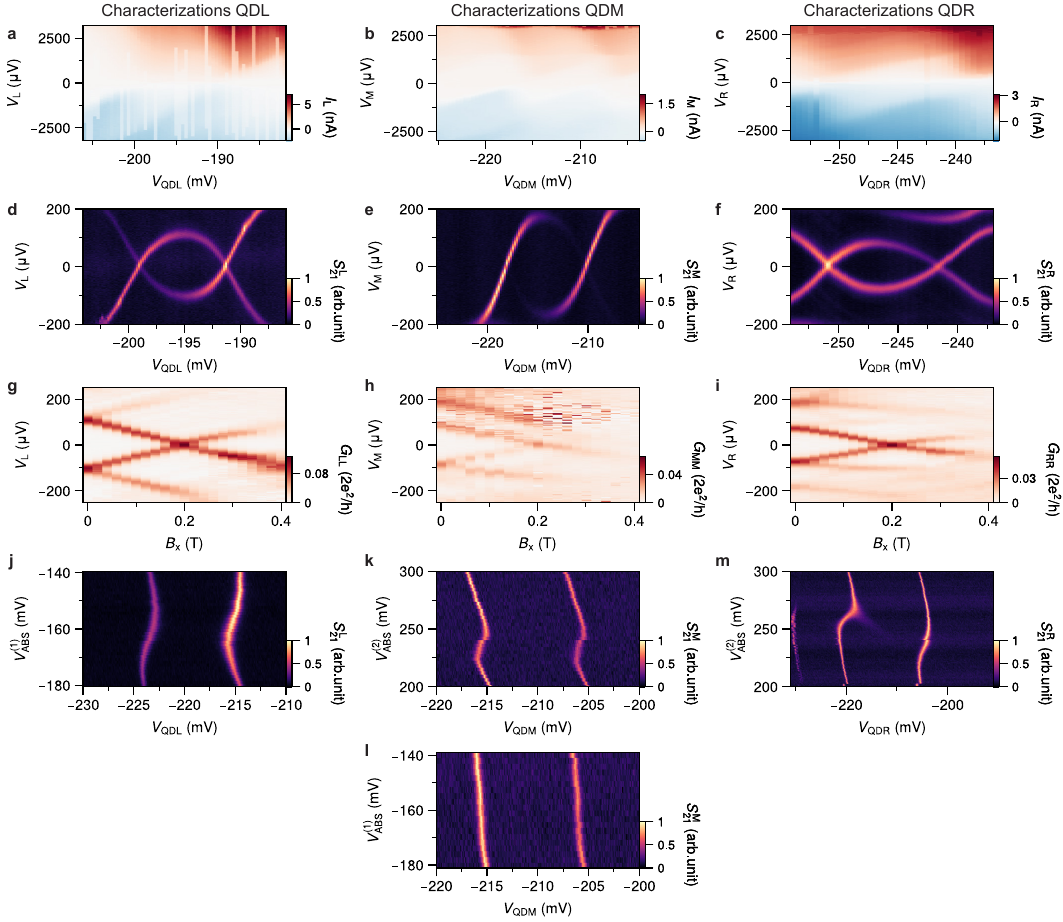}
\caption{\textbf{Quantum dot characterisations.} Results presented in the main text are obtained using a single orbital in each QD.
Characterisation measurements of each QD is shown here. 
Left, middle and right columns pertain to the left, middle and right QD respectively. 
\fl{a-c} Coulomb diamonds measured at \var{B}{x}=\SI{0}{\milli \tesla}. 
To achieve strong interdot coupling, barriers between QD and neighbouring regions are kept relatively open, such that a finite current can be observed within each Coulomb diamond. 
The outline roughly indicates the charging energies to be $>$\SI{1}{\milli\volt}.
\fl{d-f} RF-Spectroscopy of each QD in the same regime as (a-c), for a smaller range of applied voltage biases. 
In this strong coupling regime, so-called Yu-Shiba-Rusinov states form at sub-SC gap energy scales whose energies are non-linearly dependent on the plunger gate voltages~\cite{Nygaard2016}.
\fl{g-i} Spectroscopy as a function of magnetic field \var{B}{x} applied along the 1-D channel, with each plunger gate set close to the zero field charge degeneracy point based on (d-f). 
The slope of the splitting sub-gap states provides an estimate of the \var{g}{}-factors for these parameters to be 18.4, 14.8 and 13.7 respectively.
\fl{j-m} Examples of QD-ABS charge stability diagrams, used to calibrate the interdot interactions following the procedure detailed in \cite{Zatelli2023, Liu2024Enhancing}.
The discrepancies in the plunger gate voltages used for each orbital between (a-c) and (j-m) arise due to gate-jumps and cross-talk between neighbouring barrier gates.
}
\label{supp:CharactQDs}
\end{figure}

\begin{figure}[t!]
\centering
\includegraphics[width=\textwidth]{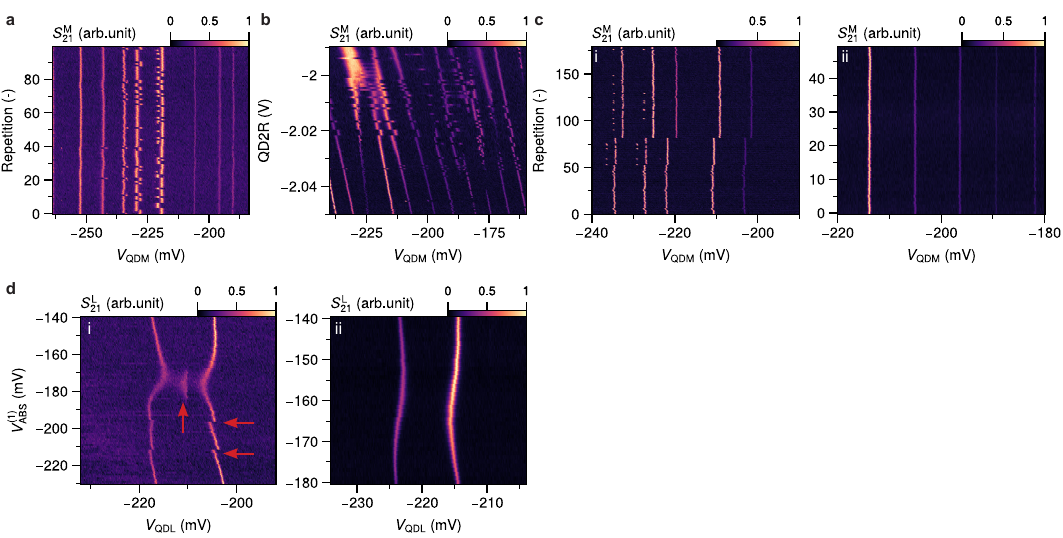}
\caption{\textbf{QD stability.} 
\fl{a} Example of repeated RF-measurements of the middle QD resonances. The same measurement was repeated 100 times, once every second. The plot shows a stack of all repetitions. 
The resonances at \SI{-225}{\milli\volt} are observed to switch between two states during this time, which makes them unsuitable for slow DC measurements.
\fl{b} In order to find a stable regime, we observe the effect of varying the barrier gates forming the QDs. 
Here, for example, the middle QD resonances are found to be more stable when the right barrier gate is tuned below \SI{-2.02}{\volt}. 
\fl{c} This becomes apparent when comparing repeated RF measurements for (i) unstable and (ii) stable regimes. 
\fl{d} In addition, charge-jumps in the gate voltages can affect the charge-stability diagrams. Panel (i) shows an example of a CSD for QDL and the left hybrid gate, where charge jumps occur in both \ABS{1} and \var{V}{QDL} (indicated by the red arrows). By fine-tuning the barrier gates forming QDL, we can reach a state where these jumps are avoided in the region of interest, shown in panel (ii). 
}
\label{supp:S3stability}
\end{figure}

\begin{figure}[t!]
\centering
\includegraphics[width=\textwidth]{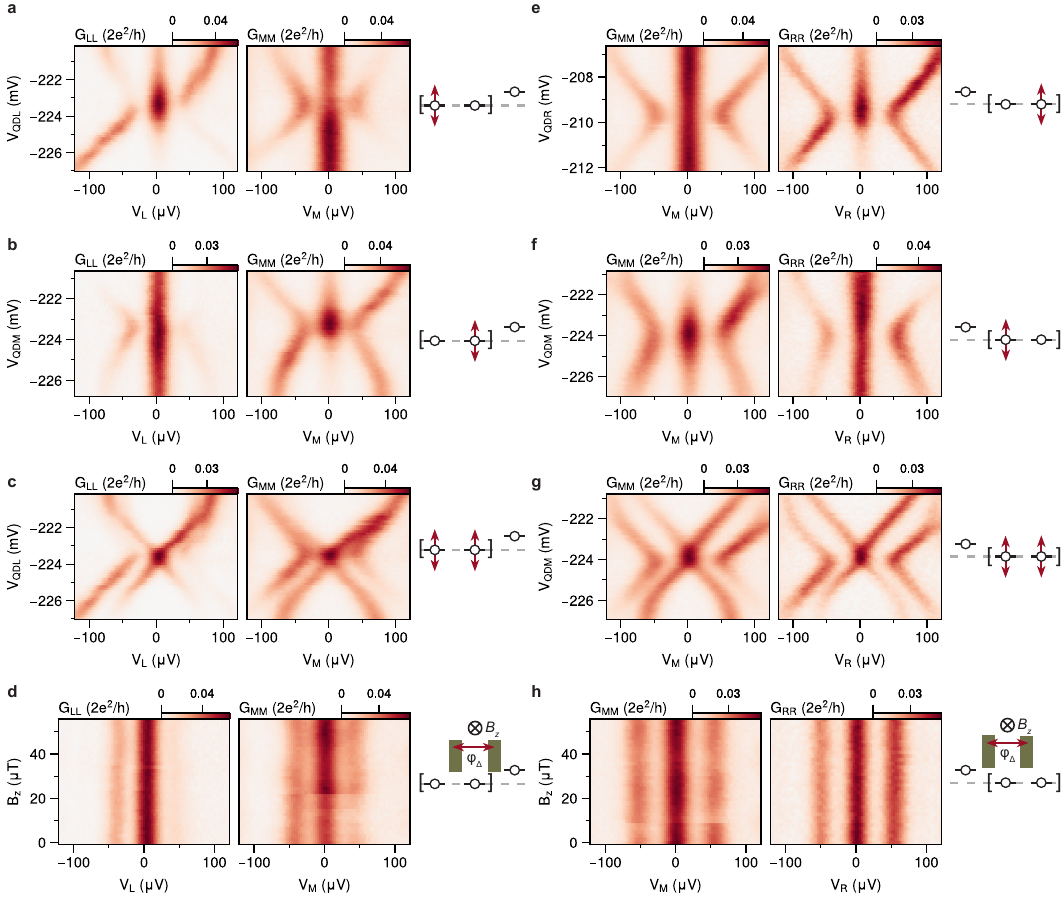}
\caption{\textbf{Conductance spectra of two-site QD pairs.} 
Studying CSDs for pairs of QDs provides information about the interdot couplings and allows one to reach the sweet spot conditions shown in Fig.~1c,d. 
In addition, finite bias conductance spectra at the two-site sweet spots should be obtained~\cite{Bozkurt2024} for all possible combinations of QD detunings.
This is shown here for the charge configuration in Fig.~1.
Adjacent schematics represent the configurations of the QDs and the parameters varied. 
Left column: tunneling spectroscopies of the left QD pair at a sweet spot (i.e. $t_{1}$~=~$\Delta_{1}$), with the right QD kept in Coulomb blockade.
\var{G}{LL} and \var{G}{MM} are measured as a function of \fl{a} detuning \var{V}{QDL}, \fl{b} detuning \var{V}{QDM}, \fl{c} detuning both \var{V}{QDL} and \var{V}{QDM} simultaneously and \fl{d} Applying a magnetic field \var{B}{z} perpendicular to the superconducting loop.
Right column: tunneling spectroscopy of the right pair of QDs at a sweet spot (i.e. $t_{2}$~=~$\Delta_{2}$), with the left QD kept in Coulomb blockade.
\var{G}{MM} and \var{G}{RR} are measured as a function of \fl{e} detuning \var{V}{QDR} \fl{f} detuning \var{V}{QDM}, \fl{g} detuning both \var{V}{QDM} and \var{V}{QDR} and \fl{h} Applying a magnetic field \var{B}{z} perpendicular to the superconducting loop.
It is important to note that (d) and (h) are obtained for the same configuration as the flux-dependence measurements in Fig.~2, with the only difference being that here the outer QD is kept in Coulomb blockade.
The lack of response to \var{B}{z} rules out more trivial origins of the flux dependence in Fig.~2, such as oscillations of the middle QDs electrochemical potential energy.
}
\label{supp:PMMspectra}
\end{figure}

\begin{figure}[t!]
\centering
\includegraphics[width=\textwidth]{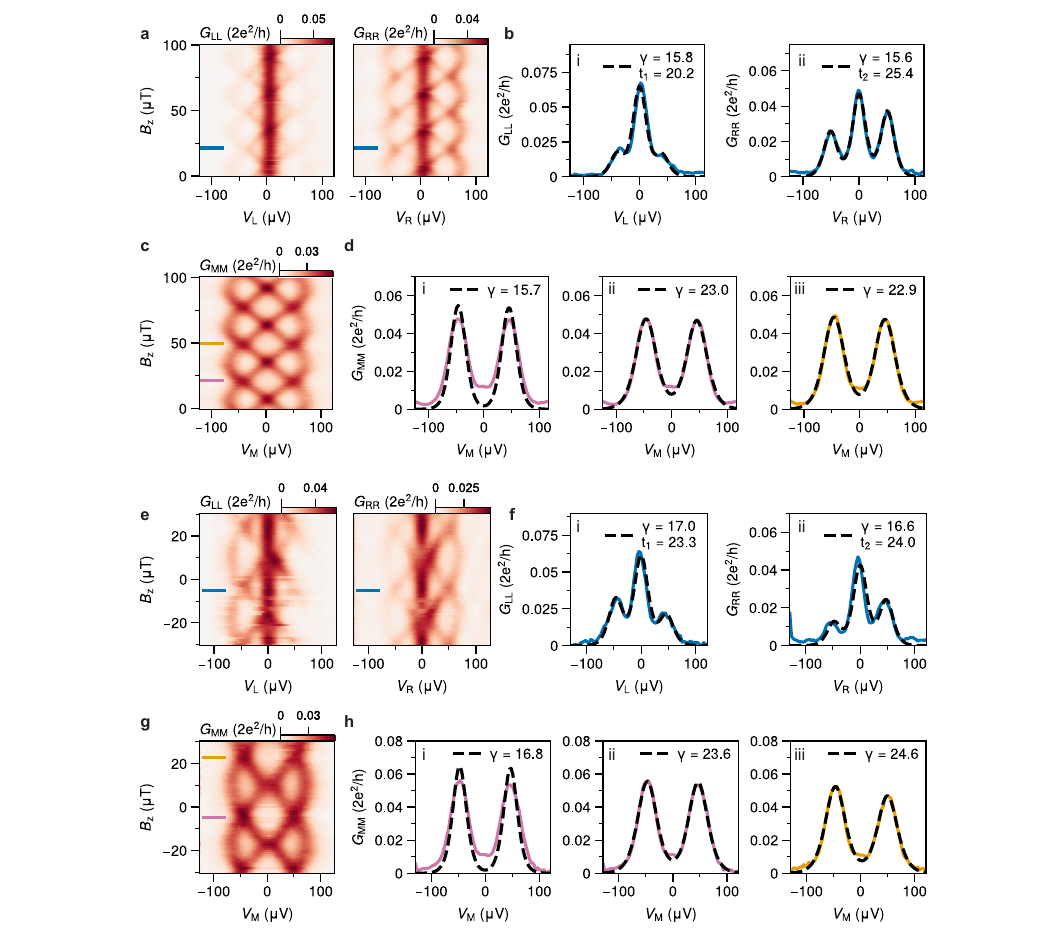}
\caption{\textbf{Fitting of conductance line-traces.} The measurements in Fig.~2 show a finite in-gap conductance in \var{G}{MM} when \var{\phi}{\Delta}~=~0. 
Possible origins for this include thermal broadening, and small deviations in the QD plunger gate voltages from being precisely at $\mu_i=0$. 
Here we address the former, by comparing the conductance bias-traces with the theoretically expected shapes in the small lead-QD coupling limit ($\Gamma \ll \mathrm{k}_{\mathrm{B}}T$) described in Methods.
\fl{a} Repetitions of conductance spectra for \var{G}{LL} and \var{G}{RR} as a function of \var{B}{z}, as shown in Fig.~2a. 
\fl{b} Line-cuts taken from the indicated position in (a), at \var{B}{z} corresponding to \var{\phi}{\Delta}~=~0.
\var{G}{LL} is fitted to Eq.~\ref{eq:Gshape_GLL} and \var{G}{RR} is fitted to Eq.~\ref{eq:Gshape_GRR}, yielding estimates for \var{t}{1} and \var{t}{2}. 
We find the conductances are well described by the temperature-limited fits and find both are described by the same broadening parameter ($\gamma$~=~15.6).
\fl{c} Repetition of conductance spectra for \var{G}{MM} as a function of \var{B}{z} from Fig.~2a. 
\fl{d} Using the \var{t}{1} and \var{t}{2} values extracted in (b), we fit the indicated \var{G}{MM} line-cut from (c) to Eq.~\ref{eq:Gshape_GMM}. In (i), $\gamma$ is fixed to be the same value as extracted in (b), while in (ii) $\gamma$ is included as fitting parameter. 
(iii) uses the same fitting as (ii), for a line-cut taken at a different $2\pi$ period as indicated in (c).
The conductance is again well-described by the temperature-limited fit.
A larger broadening parameter is however required, the origin of which is unclear and not captured by the numerical simulations.
In (ii) and (iii) the measured conductance at \var{V}{M}~=~0 is $\approx$~3$\mathrm{m}\mathrm{G}_{0}$ larger than explained by the fits. 
From Eqs.~\ref{eq:detuning}, this remainder would correspond to offsets in $\mu_{\mathrm{L}}$/$\mu_{\mathrm{R}}$ on the order of~$\pm$\SI{5}{\micro\electronvolt}.
The plunger gate voltages are set with a resolution of \SI{60}{\micro\volt}, which combined with a QD leverarm of $\approx$0.05 would translate to potential offsets on the order of \SI{3}{\micro\electronvolt} in $\mu_{\mathrm{L}}$ and $\mu_{\mathrm{R}}$.
\fl{e-h} shows a repetition of the outlined procedure, for measurement using the charge configuration shown in Fig.~\ref{supp:reprod}. 
A similar behaviour is observed.
}
\label{fig:bulkgapfit}
\end{figure}

\begin{figure}[h!]
\centering
\includegraphics[width=\textwidth]{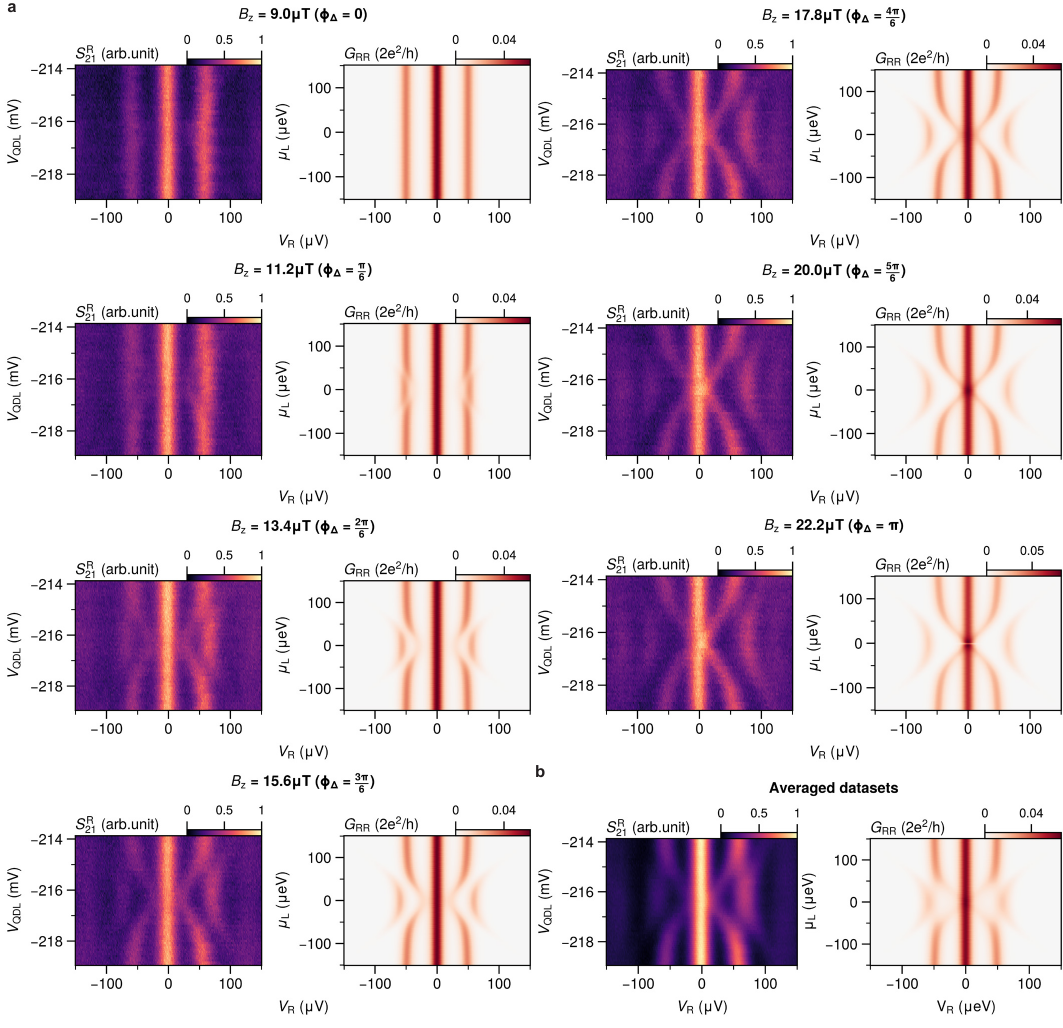}
\caption{\textbf{Conductance spectra for $\phi_{\Delta}$ between 0 and $\pi$.} In the main text, Fig.~3 shows measurements of conductance spectra at a three-site sweet spot, obtained at two $\mathrm{B}_{\mathrm{z}}$ values corresponding to $\phi_{\Delta}=0$ and $\phi_{\Delta}=\pi$. 
The ability to control the flux allows us to perform such measurements at any intermediate value of $\phi_{\Delta}$. 
\fl{a} RF-spectroscopy measurements of $\mathrm{S}_{21}^{\mathrm{R}}$ and corresponding numerical simulations, upon detuning $\mathrm{V}_{\mathrm{QDL}}$, for seven values of $\mathrm{B}_{\mathrm{z}}$ corresponding to \var{\phi}{\Delta}~=~0 and \var{\phi}{\Delta}~=~$\pi$.
The experimental evolution corresponds well to the numerical simulation at each stage.
Recent work on a three-site Kitaev chain with two separately grounded SCs~\cite{Bordin2024threesite} concludes that a small voltage difference between the SCs (on the order of \SI{}{\micro\volt}), may give rise to rapid phase oscillations. 
Hence, it is assumed their (slow) measurements reflect an average over many periods of \var{\phi}{\Delta}.
\fl{b} Shows averaged RF-spectroscopy for 25 measurements in the same range as (a), and the corresponding numerically averaged simulation, finding good agreement with the reported behaviour in~\cite{Bordin2024threesite}.
}
\label{supp:fluxaveraging}
\end{figure}

\begin{figure}[h!]
\centering
\includegraphics[width=\textwidth]{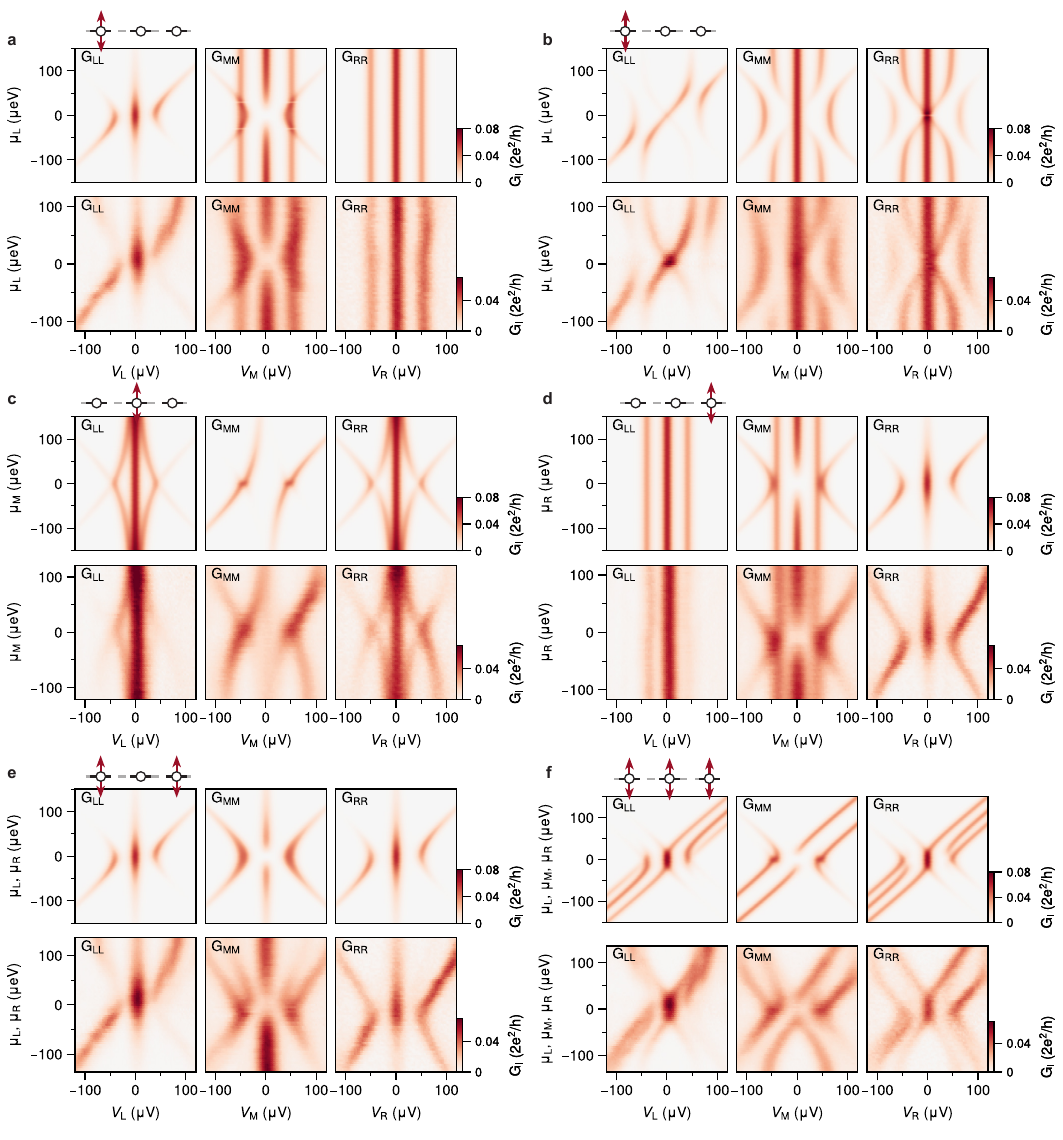}
\caption{\textbf{Three-site conductance spectra for different QD detuning combinations}.  
In a three-site Kitaev chain, the ZBPs arising on the outer QDs are expected to persist when detuning either a single QD or pairs of QDs~\cite{Bordin2024threesite}. 
Fig. 3 only demonstrates the response to detuning \var{V}{QDL}.
Here, four other possible detuning combinations are highlighted, for the same device configuration. 
Accompanying numerical simulations use the same set of parameters as shown in the main text.
Plunger gate voltages have been converted to chemical potential energies using the QD leverarms.
\fl{a} and \fl{b} show repetitions of the conductance spectra shown in Fig.~3, where \var{\phi}{\Delta}~=~0 and \var{\phi}{\Delta}~=~$\pi$ respectively, including a comparison to the numerical simulation.
Furthermore, we show the response at \var{\phi}{\Delta}~=~0 to \fl{c} Detuning \var{V}{QDM} (\var{\mu}{M}), \fl{d} detuning \var{V}{QDR} (\var{\mu}{R}), \fl{e} detuning both \var{V}{QDL} and \var{V}{QDR} and \fl{f} detuning all three QDs simultaneously.
When \var{\phi}{\Delta}~=~0, the ZBP measured in \var{G}{LL} and \var{G}{RR} only splits from zero-energy when all three QDs are detuned.
}
\label{supp:extradetuning}
\end{figure}

\begin{figure}[h!]
\centering
\includegraphics[width=\textwidth]{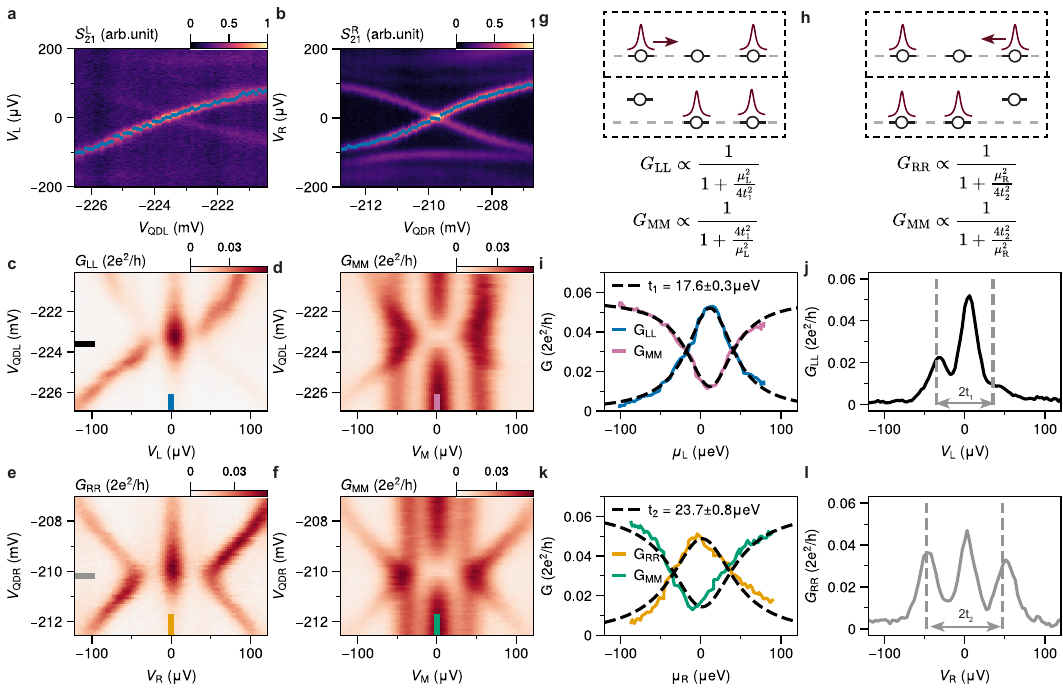}
\caption{\textbf{Shifting the MBS wavefunction - comparison to analytical result.} At a three-site Kitaev chain sweet spot, with $\phi_{\mathrm{\Delta}}=0$, detuning either of the outer QDs shifts the MBS wave-function to the middle QD. 
In Methods, we derive that this reflects in the zero-bias conductance of each site and depends only on the coupling parameters $t_{1}/t_{2}$~(=~$\Delta_1/\Delta_2$) (see Eqs.~\ref{eq:detuning}). 
To do this analysis experimentally, first the chemical-potential energies $\mu_{L}$, $\mu_{R}$ of QDL and QDR are measured as a function of \fl{a} \var{V}{QDL} and \fl{b} \var{V}{QDR}, by measuring each QD spectrum with the unused QDs in Coulomb blockade.
With all parameters tuned to the sweet spot values, we detune \var{V}{QDL} around charge degeneracy and measure \fl{c} \var{G}{LL} and \fl{d} \var{G}{MM}, as shown in Fig.~3a. 
Additionally, we detune \var{V}{QDR} and measure \fl{e} \var{G}{RR} and \fl{f} \var{G}{MM}.
As visualised in \fl{g} and \fl{h}, these experiments result in the shifting of the MBS wave-function from the outer QD to the inner QD.
In (c,d), the conductances measured at \var{V}{L},\var{V}{M}~=~0 depend only on \var{\mu}{L} and \var{t}{1}.
Similarly for (e,f) they scale according to \var{\mu}{R} and \var{t}{2}.
\fl{i} We extract \var{G}{LL} and \var{G}{MM} along \var{V}{L},\var{V}{M}~=
~0 from (c) and (d) and convert \var{V}{QDL} to \var{\mu}{L} using (a). 
Fitting the analytical formulas shown in (g), with an additional scaling factor, an estimate for \var{t}{1} of \SI{17.6}{\micro\electronvolt} is obtained.
This can be compared to the width of the excitation gap at \var{\mu}{L}~=~0, which theory predicts to be $2$\var{t}{1}. \fl{j} shows the line-trace, with the dashed lines indicating the expected location of the excited states based on the extraction in (i).
\fl{k} We repeat this procedure for \var{G}{RR} and \var{G}{MM} along \var{V}{R},\var{V}{M}~=~0 from (e) and (f), converting \var{V}{QDL} to \var{\mu}{R} using (b).
Now fitting the formulates shown in (h), we can estimate \var{t}{2} of \SI{23.7}{\micro\electronvolt}.
\fl{l} This again agrees with the excitation gap at \var{\mu}{R}~=~0.
}
\label{supp:detuningvsanalytics}
\end{figure}

\begin{figure}[h!]
\centering
\includegraphics[width=0.9\textwidth]{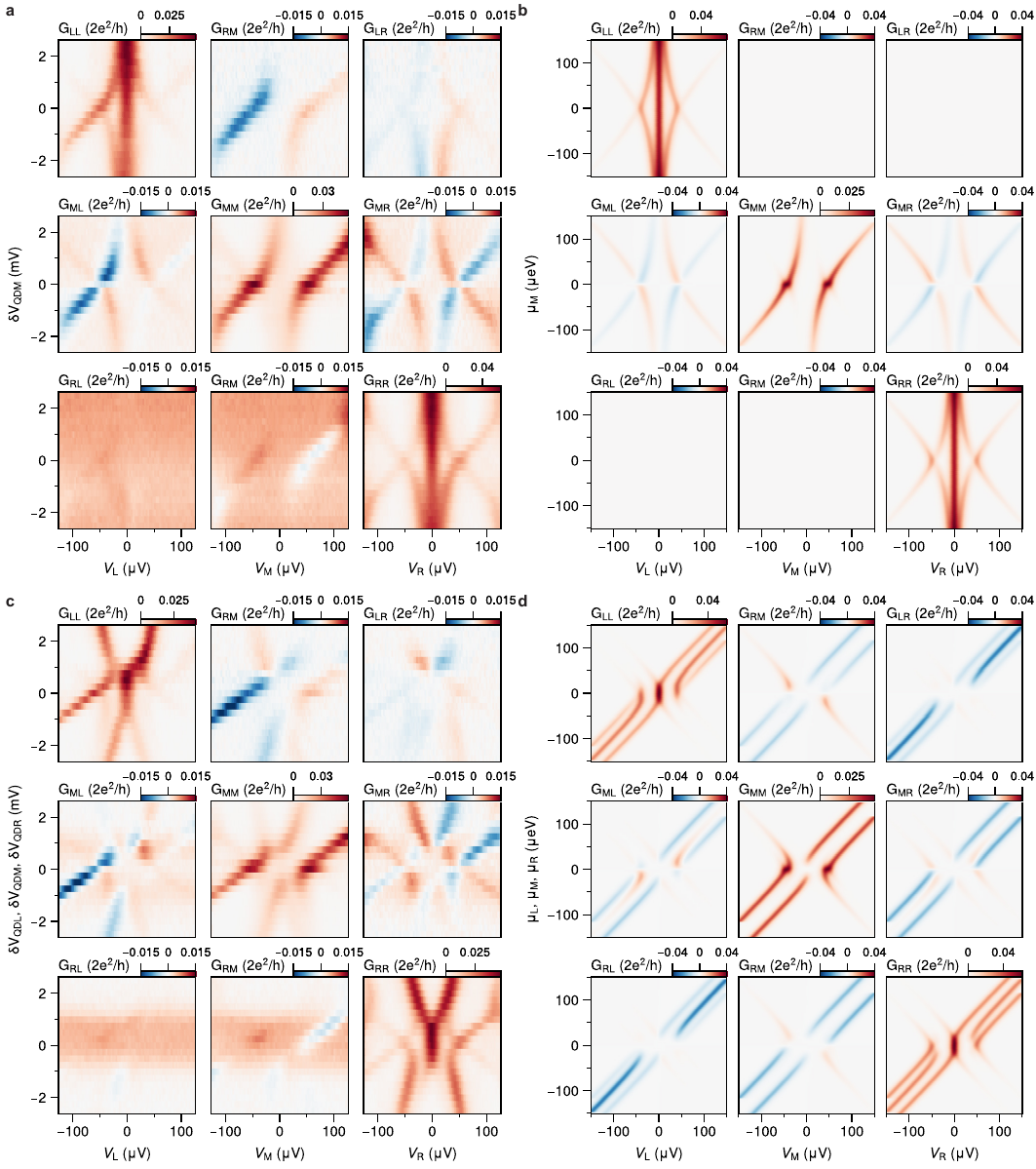}
\caption{\textbf{Full local and non-local conductance matrix measurements}. 
The main text focuses on measurements of local conductances probed through each of the three normal contacts. 
For such measurements, the non-local responses are recorded in addition, not shown due to size constraints. 
Here two measurements are highlighted, for the charge configuration shown in Fig~S4g. 
\fl{a} Local and non-local conductances when detuning \var{V}{QDM}, compared to \fl{b} numerical simulations of detuning \var{\mu}{M}.
Strikingly, the ZBPs do not appear in the non-local measurements, as expected due to arising from MBSs localised on the outer QDs.
Additionally, the same patterns of positive and negative non-local conductance in \var{G}{ML} and \var{G}{LM} are observed, with the sign inverting at $\mu$~=~0.
Unlike the simulations, signals appear in \var{G}{RM}, \var{G}{LR}, \var{G}{RL} and \var{G}{RM} that are not captured by the effective model. 
We note that a more complete model incorporating explicitly the hybrid regions such as in~\cite{Tsintzis2022} may be needed to fully describe all non-local effects.
\fl{c} Local and non-local conductances when simultaneously detuning \var{V}{QDL}, \var{V}{QDM} and \var{V}{QDR}, compared to \fl{d} numerical simulations of detuning all \var{\mu}{i}.
}
\label{supp:fullMatrix}
\end{figure}

\begin{figure}[h!]
\centering
\includegraphics[width=\textwidth]{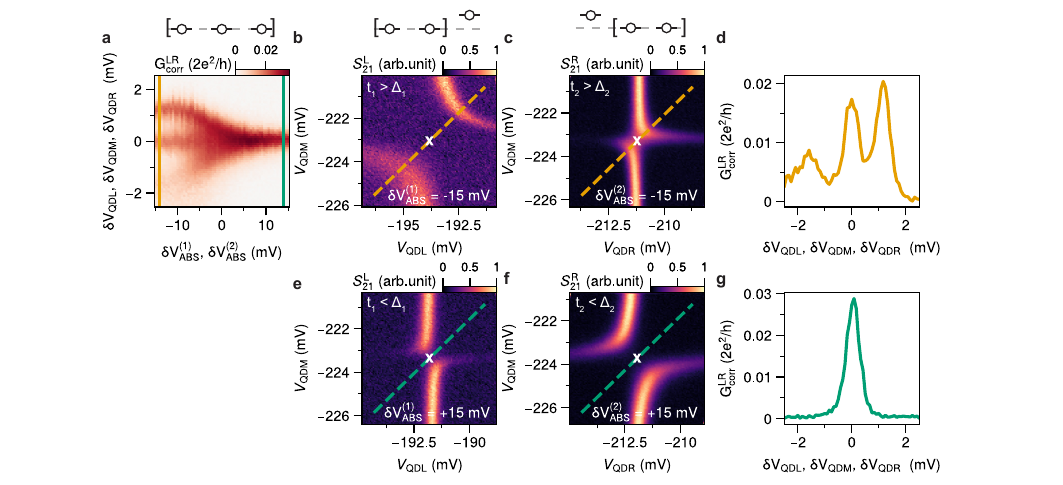}
\caption{\textbf{Details of measurement procedure for Fig.~4} In the main text, Fig.~4 highlights zero-bias conductance measurements at a three-site sweet spot, when simultaneously varying \var{V}{QDL}, \var{V}{QDM} and \var{V}{QDR} against simultaneously sweeping \ABS{(1)} and \ABS{(2)}.
Here, we detail how these measurements are performed. 
\fl{a} shows a repetition of the data as shown in Fig.~4e.
First, the sweet spot values for \ABS{(1)} and \ABS{(2)} were determined from measuring CSDs, through the process shown in Fig.~1. 
We denote these values $\delta$\ABS{(1)}=0 and $\delta$\ABS{(2)}=0 respectively.
Next, we apply \SI{15}{\milli\volt} to each, with respect to these sweep spot voltages, in the direction that results in an avoided crossing signifying $t>\Delta$.
Due to cross-coupling, the resonance value for each QD needs to be re-determined. 
This is done by measuring the zero-bias Coulomb resonance for each QD, with the two other QDs set off resonance, and fitting a Lorentzian line-shape to determine the centre. 
\fl{b} and \fl{c} show CSDs measured at these $V_{\mathrm{ABS}}$ values for the left-middle and middle-right pairs respectively, taken for verification of the interdot coupling.
Both sides show an antidiagonal avoided crossing signifying \var{t}{i}$>$\var{\Delta}{i}. 
The centers as obtained by the centering procedure are marked by the crosses.
After centering, all three QDs are brought on resonance and swept simultaneously, accounting for differences in leverarms (see Fig.~\ref{supp:detuningvsanalytics}).
Here the orange line-trace in (a) is obtained, shown in \fl{d}.
The same procedure is repeated for every  set-point, until reaching \ABS{(1)}, \ABS{(2)}~=~$+$\SI{15}{mV}.
Now both CSDs show a clear diagonal avoided crossing, shown in \fl{e} and \fl{f}. 
The new centers are marked, differing within a few mV from those in (b)/(c). 
The green dashed lines mark the paths taken by the simultaneous sweep for the green line in (a), plotted in \fl{g}.
}
\label{supp:ABSQDCSD}
\end{figure}

\begin{figure}[h!]
\centering
\includegraphics[width=\textwidth]{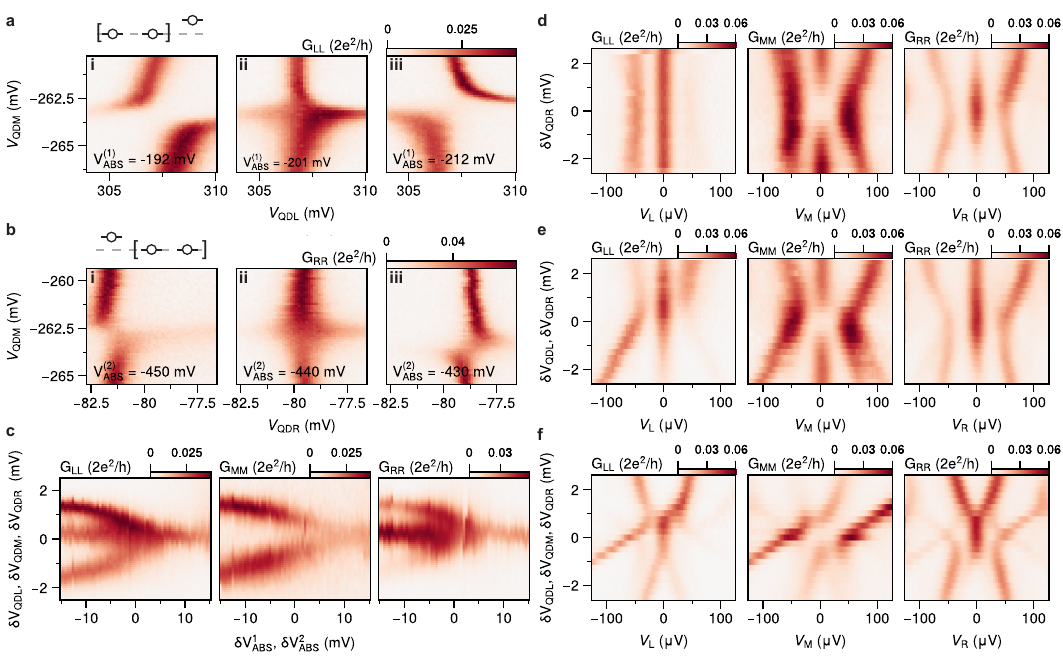}
\caption{\textbf{Reproduction of main results in a separate cooldown.}
The measurements in Fig.~4 were repeated for validation during a separate cooldown of the same device. 
First, two-site sweet spots were obtained for both two-site pairs, as in Fig.~1. 
\fl{a} CSDs for resonances in the left and middle QD, varying \ABS{1} in a range where the avoided crossing changes direction to determine the sweet spot (panel ii). Similarly, a sweet spot was obtained for the middle and right QD, upon varying \ABS{2}, shown in \fl{b}. The magnetic field \var{B}{z} corresponding to \phiD~=~0 was determined from the spectroscopy measurements shown in Fig.~\ref{fig:bulkgapfit}g. \fl{c} In this configuration, the measurement shown in Fig.~4 was repeated, using the same procedure as detailed in Fig.~\ref{supp:ABSQDCSD}. Additionally, we reproduce here the conductance spectra at the sweet spot as a function of \fl{d} detuning \var{V}{QDR}, \fl{e} detuning both \var{V}{QDL} and \var{V}{QDR} and \fl{f} detuning simultaneously all 3 QDs.
}
\label{supp:reprod}
\end{figure}

\clearpage
\bibliography{ref}
